\documentclass[aps,prd,showpacs,showkeys,floatfix,nofootinbib,onecolumn, 
notitlepage,superscriptaddress]{revtex4-1}
\usepackage{natbib}
\usepackage{color}
\usepackage{graphicx}
\usepackage{epsfig}
\usepackage{bm}
\usepackage{amsthm}
\usepackage{amsfonts}
\usepackage{float}
\usepackage{amsmath,amssymb}
\usepackage{graphicx}
\usepackage[caption = false]{subfig}
\usepackage{hyperref}
\usepackage{appendix}
\begin{document}
\title{Chiral condensate from a hadron resonance gas model  }
\author{Deeptak Biswas}
\email{deeptakb@imsc.res.in}
\affiliation{The Institute of Mathematical Sciences, a CI of 
Homi Bhabha National Institute, Chennai, 600113, India}
\author{Peter Petreczky}
\email{petreczk@bnl.gov}
\affiliation{Physics Department, Brookhaven National Laboratory, Upton 
NY 11973, USA}
\author{Sayantan Sharma}
\email{sayantans@imsc.res.in}
\affiliation{The Institute of Mathematical Sciences, a CI of 
Homi Bhabha National Institute, Chennai,  600113, India}
%
%
\begin{abstract}
In this work we address the question of how well the chiral
crossover transition can be understood in terms of a noninteracting
hadron resonance gas model. Using the latest results on the variation of
hadron masses as a function of the pion mass from lattice quantum
chromodynamics, we study the temperature dependence of the renormalized
chiral condensate in 2+1 flavor QCD. Furthermore, we suggest a better 
criterion to estimate of the pseudocritical temperature, which gives  
$T_c = 161.2 \pm 1.7$ MeV, which is much improved compared to all the 
earlier results within the hadron resonance gas model or chiral 
perturbation theory. For the curvature of the pseudocritical line we 
find $\kappa_2 = 0.0203(7)$, which is in very good agreement with 
continuum extrapolated lattice results.
\end{abstract}
\maketitle
%
\section{Introduction}
The breaking of chiral symmetry is one of the key features of quantum 
chromodynamics (QCD) underlying much of our understanding of hadron physics. 
The breaking of this symmetry is signaled by the nonzero value of the quark (chiral) 
condensate, $\langle \bar \psi \psi \rangle \ne 0$. It is well known that 
at high temperature the chiral symmetry is restored and this restoration 
happens as an analytic crossover at a temperature $T_c=156.5 \pm 1.5$ 
MeV~\cite{HotQCD:2018pds}. The chiral condensate rapidly decreases at and 
above $T_c$. Below the crossover temperature, it is expected that 
the thermodynamics of QCD matter can be understood in terms of hadronic 
degrees of freedom. There is a wide belief that the description of 
QCD matter as a gas of hadrons is valid up to the crossover temperature, 
$T_c$. This belief is the basis of our current understanding of the 
freeze-out condition in heavy-ion collisions~\cite{Becattini:2005xt, 
Andronic:2005yp, Andronic:2008gu, Manninen:2008mg, Bhattacharyya:2019wag, 
Bhattacharyya:2019cer, Biswas:2020dsc}. These considerations also 
implicitly assume that for $T<T_c$ the decrease in the chiral condensate 
can be understood in terms of a hadron gas. Therefore, an important 
question to clarify is to what extent we can understand the decrease of 
the chiral condensate for $T<T_c$ in terms of hadronic degrees of 
freedom. This question is also important for our understanding of the 
QCD phase diagram at large values of baryon density. Recent studies 
suggest that lattice QCD calculations based on Taylor expansion in 
chemical potential or analytic continuation may only work for baryon 
chemical potential $\mu_B<2.5~T$,~\cite{Bazavov:2017dus, 
Borsanyi:2018grb, DElia:2016jqh,Bollweg:2022rps} though another study 
claims that lattice QCD calculations may be useful for baryon chemical 
potential as high as $3.5~T$~\cite{Borsanyi:2021sxv}. In any case, the 
study of the temperature dependence of the chiral condensate at large 
values of $\mu_B$ is of great interest and can be performed within the 
framework of a hadron gas.

The hadron resonance gas (HRG) model  turned out to be very successful in 
describing QCD thermodynamics below the crossover temperature. The main 
idea of the HRG model is that the interacting gas of hadrons can be 
replaced by a noninteracting gas of hadron and hadron resonances, i.e., all 
interactions between hadrons can be approximated by the creation of additional 
hadronic resonances. This approach can be justified in terms of 
relativistic virial expansion, where interactions between hadrons are 
expressed in terms of phase shifts~\cite{Dashen:1969ep}. Using 
experimentally known phase shifts within this approach, it has been shown 
that the nonresonant part of the phase shifts largely cancel out in the QCD 
pressure, and indeed the interacting part of the pressure can be well 
described by the contribution of resonances treated as stable hadrons~
\cite{Venugopalan:1992hy}.

There has been an extensive effort to test the validity of the HRG model by 
comparison to the lattice QCD results. Early works in this direction were presented in Refs.~\cite{Karsch:2003vd,Karsch:2003zq,Ejiri:2005wq,Huovinen:2009yb} and confirmed the validity of HRG approach. Recent continuum extrapolated lattice QCD results on the equation of state show that HRG works reasonably well below $T_c$ if all known resonances from the Particle Data Group (PDG) are included~\cite{Borsanyi:2010bp, Bazavov:2012vg, Bazavov:2014pvz}. A precision lattice study of the second derivatives of the pressure with respect to the chemical potentials again confirmed the validity of the HRG model~\cite{Bollweg:2021vqf}. However for the quantities involving strangeness, for example, 
strangeness-baryon-number correlations, including additional resonances not listed by the PDG but predicted by QCD (or QCD inspired models) turned out to be necessary~\cite{Bollweg:2021vqf,Bazavov:2014xya}. The need for additional resonances to describe the baryon-strangeness correlations was also confirmed using relativistic virial expansion with state-of-the-art partial wave analysis of the phase shifts 
\cite{Fernandez-Ramirez:2018vzu}. Furthermore, the baryon-charm
correlation calculated on the lattice requires taking into account 
resonances not listed by the PDG~\cite{Bazavov:2014yba}. For higher-order 
derivatives of the pressure with respect to the chemical potential the 
HRG model does not work as well. It has been argued that repulsive 
baryon-baryon interactions play an important role in this case and the 
contribution of these repulsive interactions increases with the order of 
the derivative. Including these interactions in some phenomenological way 
improves the agreement between the lattice results and HRG 
model~\cite{Vovchenko:2016rkn,Huovinen:2017ogf,Vovchenko:2017drx,Vovchenko:2017gkg}. 

The temperature dependence of the chiral condensate has also been 
studied in the HRG model~\cite{Toublan:2004ks,Tawfik:2005qh,Borsanyi:2010bp,Jankowski:2012ms,Andersen:2022clu} and within the chiral perturbation theory~\cite{GERBER1989387,GarciaMartin:2006jj}. These calculations show a significant change in the chiral condensate in the vicinity of the crossover temperature. The pseudocritical temperature extracted from the point where the chiral condensate of a pion gas within the next-to-next-to-leading order (NNLO) chiral perturbation theory falls to zero is about $\gtrsim 250$ MeV, 
which is lowered to about $190$ MeV when additional hadrons are included~\cite{GERBER1989387}. However, the uncertainty in the HRG calculations of the chiral condensate is difficult to quantify. This is due to the fact that, in order to obtain the chiral condensate in the HRG model one has to know the precise dependence of the hadron masses on the light quark mass. Except for the few lowest-lying hadron states this dependence is not well known. The main goal of this paper is to revisit the temperature dependence of the chiral condensate in the HRG model and quantify the uncertainties related to the poorly known quark mass dependence of hadron masses. We will show that with systematic improvements in the procedure we can make a much more precise estimate of the $T_c$ within the HRG model.

The paper is organized as follows: In the next section, we discuss 
the general formalism and the relevant observables. In particular, 
we focus on two renormalized definitions of subtracted chiral condensate, 
which we calculate within the HRG model. We show how much contribution 
to these observables comes from different ground state mesons and baryons as well 
as their higher excited states in Sec. III. We find that the 
dominant contribution to the renormalized chiral contribution at small 
baryon densities comes from the mesons, in particular, the ground state 
pseudoscalar and vector mesons. We have extracted the pseudocritical 
temperature $T_c=161.2\pm1.7$ MeV and a curvature of the pseudocritical line $\kappa_2=0.0203(7)$ corresponding to the chiral crossover transition which not only is in very good agreement with the latest lattice QCD result but can be obtained with considerably less computational effort. Furthermore, our estimates of the pseudocritical temperature are much improved compared to the corresponding calculations within the chiral perturbation theory. Though pseudoscalars dominantly contribute to the chiral condensate, extending this formalism to the chiral limit has its own limitations. We discuss this in detail, showing how we could improve from the previous works in Ref.~\cite{GERBER1989387}. We conclude by discussing the important implications of our work, in particular, the possible extension of these calculations to understand the QCD phase diagram at finite baryon densities where lattice QCD calculations are severely limited due to the infamous \emph{sign problem}; see e.g. Ref.~\cite{Schmidt:2017bjt} 
for a recent review.

\section{Chiral Condensate in the Hadron Resonance Gas model}
\subsection{Renormalized definitions of chiral condensate}
\label{subsec:introhrg}

The total pressure due to a noninteracting ensemble of several species 
of hadron labeled by $\alpha$, enclosed within a volume $V$ and at an 
ambient temperature $T$, is given as,
\begin{equation}
P=\pm \sum_\alpha \frac{g_{\alpha} T}{2 \pi^{2}} 
\int_{0}^{\infty} p^{2} d p ~
\ln \left[1 \pm \exp \left(-\left(E_{\alpha}-\mu_{\alpha}\right) / T\right)\right]~,
\label{Eq.P_HRG}
\end{equation}
Here, $+(-)$ corresponds to baryons (mesons) respectively. The degeneracy 
factor and the energy of each particle $\alpha$ are denoted by
$g_{\alpha}$ and $E_{\alpha}$, respectively. Furthermore, 
$E_\alpha=\sqrt{p^2+M_\alpha^2}$ with $p, M_{\alpha}$ being the 
magnitude of the particle momentum and its mass respectively. 
We consider here a grand canonical ensemble with each species 
of hadron denoted by a chemical potential $\mu_\alpha = B_\alpha\mu_B+Q_\alpha\mu_Q+S_\alpha\mu_S$,  
which depends on the quantum numbers $B_\alpha$, $Q_\alpha$, and $S_\alpha$ 
denoting the conserved quantum numbers baryon number, electric charge, and 
strangeness respectively. Within this model the interactions between hadrons 
are taken care of by inclusion of hadron resonances as free stable particles
~\cite{Dashen:1969ep, Venugopalan:1992hy}, i.e., the sum in Eq. [\ref{Eq.P_HRG}] 
also includes resonances. The light quark condensate, also called the chiral condensate, at nonzero temperature is defined as
\begin{equation}
\langle\bar{\psi}\psi\rangle_{l,T}=\langle\bar{\psi}\psi\rangle_{l,0} + 
\frac{\partial P}{\partial m_l}~,
\label{eq.defppbar}
\end{equation} 
where $m_l$  is the light quark mass. We consider the case of two degenerate light quarks, $m_u=m_d=m_l$.
The zero temperature light quark condensate $\langle\bar{\psi}\psi\rangle_{l,0}$
can be thought of as the derivative of the vacuum pressure with respect to $m_l$, and it
contains all the multiplicative as well as additive divergences (for $m_l \ne 0$ case).
To have a finite observable which is also renormalization group invariant
we consider the following combination
\begin{eqnarray}
-m_s\left[\langle\bar{\psi}\psi\rangle_{l,T}
-\langle\bar{\psi}\psi\rangle_{l,0}\right]=-m_s\frac{\partial P}{\partial m_l}~,
\label{eq.sub_psibarpsi}
\end{eqnarray}
where $m_s$ is the strange quark mass. A definition of the renormalized chiral condensate that has been discussed in the literature ~\cite{Borsanyi:2010bp} constructed out of the quantity defined 
in the Eq.~[\ref{eq.sub_psibarpsi}] but with a different normalization factor, is
\begin{equation}
 \langle\bar{\psi}\psi\rangle_{R} = 
-\frac{m_l}{m_\pi^4} \left[\langle\bar{\psi}\psi\rangle_{l,T}-
\langle\bar{\psi}\psi\rangle_{l,0}\right]~.
\label{Eq.psibarpsi_R}   
\end{equation}
Another, possibly a more natural way to define a dimensionless 
chiral condensate at nonzero quark mass ~\cite{Bazavov:2011nk} is
\begin{eqnarray}
\Delta^l_R=d+ m_s r_1^4 \left[\langle\bar{\psi}\psi\rangle_{l,T}-
\langle\bar{\psi}\psi\rangle_{l,0}\right]~,
\label{Eq.relation1}
\end{eqnarray}
where the parameter $r_1$ is derived from the static quark potential~
\cite{Aubin:2004wf} and $d=r_1^4 m_s (\lim_{m_l \rightarrow 0} \langle \bar \psi \psi \rangle_{l,0})^R$. This definition takes advantage of the fact that, in the chiral limit, the light quark condensate has only a multiplicative renormalization. The superscript $R$ denotes the renormalized quantity. Taking into account that $(\lim_{m_l \rightarrow 0} \langle \bar \psi \psi\rangle_{l,0})^R = 2 \Sigma$ and using
the values of the low energy constant of SU(2) chiral perturbation theory ($\chi$PT), $\Sigma^{1/3}=272(5)$ MeV and  $m_s=92.2(1.0)$ MeV in the $\overline{\rm MS}$ scheme at $\mu=2$ GeV from the FLAG 2022 review for the 2+1 flavor case~\cite{Aoki:2021kgd}, as well as $r_1=0.3106$ fm~\cite{MILC:2010hzw}, we obtain the value of $d=0.022791$.

In our calculations we have included all hadron resonances predicted from the quark models~\cite{Capstick:1986ter, Ebert:2009ub} in addition to those mentioned in the Particle Data Group 2016 lists, as tabulated in~Ref.~\cite{Alba:2020jir}. The HRG model constructed out of this augmented hadron list is called the QMHRG, and it turned out
to be successful in describing the temperature dependence of cumulants of different conserved charges obtained from lattice QCD for $T\lesssim 150~$MeV~\cite{Alba:2017mqu, Alba:2017hhe, Alba:2020jir, Karthein:2021cmb}. Recently, the HotQCD Collaboration constructed a HRG model with a similar list of states for their recent comparative study~\cite{Bollweg:2021vqf} on the second-order fluctuations of net baryon number, strangeness, and electric 
charge and their correlations. We have also verified that the HRG  model used in this work can reproduce the results from Ref.~\cite{Bollweg:2021vqf} with a very good precision. 

It is clear that one of the most important ingredients needed for the calculation of the chiral condensate is the dependence of the hadron masses, including those of the resonances, on the light quark mass $m_l$. Although this variation is well understood only for a few primary hadrons, more specifically the ground states, we will discuss in this work the detailed systematics for the resonance states. In the following subsections, we will separately discuss the contributions of the pseudo-Goldstone modes, i.e., those of pions and kaons, and of all other heavier hadrons 
and resonances to the chiral condensate.

\subsection{Contributions from pions and kaons}  
We first consider the contribution of pions and kaons to the chiral 
condensate. The contribution of pions and kaons to $m_s 
(\partial P/\partial m_l)$ can be written as
\begin{equation}
m_s \frac{\partial P}{\partial m_l}=-\frac{m_s}{m_l} \sum_{\alpha=\pi,K} \frac{g_{\alpha}}{2 \pi^2}
\int_0^{\infty} dp~ p^2~ n_{B}~(E_{\alpha}) \frac{1}{2 E_{\alpha}} m_l \frac{\partial M_{\alpha}^2}{\partial m_l}.
\label{Eq:PSppbar}
\end{equation}
Here $n_{B}$ is the Bose-Einstein distribution.
Extensive lattice studies of the quark mass dependence of the 
pseudoscalar meson masses found excellent agreement with SU(2) $\chi$PT, 
whereas the agreement with the SU(3) $\chi$PT is not so great. For more 
comprehensive reviews, we refer to Refs.~\cite{RBC-UKQCD:2008mhs, 
MILC:2009ltw, MILC:2009riz, Durr:2013koa, 
Budapest-Marseille-Wuppertal:2013vij, Boyle:2015exm}. 
Evidence from lattice studies suggests that an extended version of the 
SU(2) $\chi$PT, considering the strange quarks as heavier degrees of freedom~\cite{Sharpe:2006pu,RBC-UKQCD:2008mhs} can be useful in 
understanding the quark mass dependence of the kaon mass and decay 
constant. We will henceforth refer to the two-flavor $\chi$PT results 
for the pion mass from Ref.~\cite{Gasser:1983yg} and its extended 
version from Ref.~\cite{RBC-UKQCD:2008mhs} for the kaon mass. The 
dependence of the pion and kaon masses on light and strange quark 
masses $m_l, m_s$ described in terms of SU(2) low energy constants $F, 
B, \bar{l}_3, F_\pi$, are given as
\begin{eqnarray}
M_\pi^2=M^2\left[1-\frac{1}{2}\zeta~\bar{l}_3+\mathcal{O}(\zeta^2)\right]~~,~~
\zeta=\frac{M^2}{16 \pi^2 F_\pi^2} \nonumber \\
M_K^2= B_K(m_s)m_s\left[1+\frac{\lambda_1(m_s)+\lambda_2(m_s)}{F^2} M^2\right]~,~M^2=2 B m_{l},~B=\Sigma/F^2.
\label{Eq.GMOR}
\end{eqnarray}
The $\chi$PT relations for the pseudoscalar mesons are for the 
two-degenerate flavor case $m_u=m_d=m_l$, where most lattice QCD calculations 
exist in the continuum limit. The bag constant $B_K(m_s)$ and the NLO low 
energy constants $\lambda_{1,2}(m_s)$ in the kaon mass can be expressed in terms of the SU(3) low energy constants $L_4,L_5,L_6,L_8,B_0,F_0$ as~\cite{RBC-UKQCD:2008mhs},
\begin{eqnarray}
\nonumber
&& \lambda_1(m_s)=\frac{1-2 \ln \left(\frac{4 B_0 m_s}{3 \Lambda_\chi^2}\right)}{72 \pi ^2}
+32 (2 L_6-L_4)~,\\\nonumber
&& \lambda_2(m_s)=\frac{F_0^2}{2 B_0 m_s}+\frac{\frac{2}{9} 
\ln \left(\frac{4 B_0 m_s}{3\Lambda_\chi^2}\right)-\frac{1}{4} 
\ln \left(\frac{4 B_0 m_s}{3\Lambda_\chi^2}\right)}{4 \pi ^2}+8 (2L_8-L_5)-32 L_6~, \\
&& B_K(m_s)=B_0\left[1+\frac{2 B_0 m_s}{F_0^2}(32L_6-16L_4+16 L_8-8 L_5)
+\frac{2 B_0 m_s}{36 \pi^2 F_0^2}\ln\left(\frac{4 B_0 m_s}{\Lambda_\chi^2}\right)\right]~.
\label{Eq.SU2Lecs}
\end{eqnarray}
We have set the scale appearing in the above formulas 
$\Lambda_\chi=770$ MeV, i.e., at the $\rho$ meson mass. We have 
taken the values of the different $\chi$PT low energy 
constants from the latest FLAG review~\cite{Aoki:2021kgd}. 
In particular we use the value $\Sigma^{1/3}=272(5)$ MeV for 2+1 
flavor QCD. The other SU(2) low energy constant $F$ is 
obtained from the relation $F_\pi/F=1.062(7)$, where 
$F_\pi=92.2(1)$ MeV. The various SU(3) low energy constants 
that appear in the the expression of the kaon mass squared for 
$2+1$ flavor QCD are $\Sigma_0^{1/3}=245(8)$ MeV, $F_0=80.3\pm 6.0$ 
MeV and $B_0=\Sigma_0/F_0^2$. Furthermore, 
$L_4=-0.02(56)\times10^{-3},~L_5=0.95(41)\times 10^{-3},
~L_6=0.01(34)\times 10^{-3},~L_8=0.43(28)\times 10^{-3}$. 
Even though these quantities are not very precisely determined 
they account for a tiny contribution to $M_K^2$ through the 
bag constant $B_K(m_s)$ and the combination 
$\lambda_1(m_s)+\lambda_2(m_s)$. For example, in the expression for 
$\lambda_2(m_s)$ the dominant term is $F_0^2/(2m_sB_0)$, and for 
$\lambda_1(m_s)$ it is the logarithmic term. Hence the errors on 
$L_4-L_8$ have little effect for the final error on $M_K^2$. 
We also took the latest value of $m_l=3.381(40)$ MeV from 
the same review. We then calculated the derivative of the 
pseudoscalar meson masses in Eq.~(\ref{Eq.GMOR}) with respect 
to $m_l$. These derivatives were evaluated at the physical point 
as well as for the vanishing light quark mass. The latter are needed 
for the subsequent discussions in Sec.~\ref{Sec:Condchirallim}.
Lattice QCD studies show that, at the physical point, 
$m_l\partial M_\pi/\partial m_l=M_\pi/2$ to a very good 
approximation~\cite{Bali:2016lvx}. This is equivalent to the assumption 
that close to the physical point, $M_\pi^2=2Bm_l$, and hence the error 
on the $m_l$ derivative of $M_{\pi}$ can be estimated by propagating 
the errors on $B$ and $m_l$ as quoted in the FLAG review.

\subsection{Contributions from heavier hadrons}
To obtain the contribution from heavier hadrons to the chiral condensate 
we can write
\begin{equation}
m_s \frac{ \partial P}{\partial m_l} =-\frac{m_s}{m_l} \sum_{\alpha}
\frac{ g_{\alpha}}{2 \pi^2}\int_0^{\infty} dp~ p^2~n_\alpha (E_{\alpha}) \frac{M_{\alpha}}{E_{\alpha}} \sigma_{\alpha},
\end{equation}
with $n_{\alpha}$ being the Bose-Einstein distribution for $\alpha\equiv\text{mesons}$ and the Fermi-Dirac distribution for 
$\alpha\equiv \text{baryons}$. Here we introduce the so-called 
$\sigma$ terms:
\begin{equation}
\sigma_{\alpha}=m_l \frac{\partial M_{\alpha}}{\partial m_l}|_{m_l=m_l^{phys}}
=m_l \langle\alpha| \bar u u+\bar d d|\alpha\rangle=M_{\pi}^2\frac{\partial 
M_{\alpha}}{\partial M_{\pi}^2}|_{M_{\pi}=M_{\pi}^{phys}}.
\label{Eq:SigmaDef}
\end{equation}
For the ground state baryons, the $\sigma$ terms are known from the fits 
of the low energy data of masses within the chiral effective theory and 
the lattice. These are summarized in Table~\ref{table:sigma}. For other 
hadrons we will use the lattice results on the pion mass dependence of 
the hadron mass for constant strange quark mass set to its physical 
value, to evaluate 
$M_{\pi}^2\frac{\partial M_{\alpha}}{\partial M_{\pi}^2}$. A summary of 
these evaluations is given in Table~\ref{table:sigmasummary} 
and the corresponding details are given in Appendix~\ref{appdx:a}.

For the ground state pseudoscalar meson $\eta$ and vector mesons 
$\rho(770)$ and $K^*(892)$, we have collected the lattice data for the 
variation of these masses as a function of the pion mass from 
Refs.~\cite{Bali:2015gji, Guo:2016zos, Bali:2021qem} and have extracted 
the corresponding $\sigma$ terms, details of which are discussed in Appendix~\ref{appdx:a}.

Apart from the vector and pseudoscalar mesons, we have categorized
the other meson-resonances in three groups according to their quantum
numbers as shown in Table~\ref{table:sigmasummary}. Contribution to the 
chiral condensate from mesons with the same quantum numbers as $\rho$ 
(like $a$, $b$, and higher excited states of $\rho$) have been included considering their $\sigma$ values similar to $\rho(770)$, whereas the 
contribution from higher excited states of $K$ and $K^*$ have been 
included considering their $\sigma$ terms to be the same as $K^*(892)$. 
The $\sigma$ terms for the isoscalar mesons and their excited states in 
our calculations are kept identical to those of the corresponding 
ground states mesons ($\eta, \omega, \phi$). We have also included the contribution of the $\eta'$  to the renormalized chiral condensate 
separately, considering its $\sigma$ term, details of which are again 
mentioned in Appendix~\ref{appdx:a}. The contribution of the scalar 
meson $f_0 (500)$ is not considered, as it was observed earlier that 
the pole due to attractive interactions between them in the $I=0$ 
sector are exactly canceled by the repulsive 
interactions~\cite{Broniowski:2015oha}.

\begin{table}[tbh!]
\caption{\label{table:sigma} Sigma terms of baryon octet and 
decuplet ground states used in this work taken from Ref.~\cite{Copeland:2021qni}.}
\begin{tabular}{|c|c|c|c|c|c|c|c|c|c|}
\hline
Baryon states & N & $\Lambda$ & $\Sigma$ & $\Xi$ & $\Delta$ & $\Sigma^*$ 
& $\Xi^*$ & $\Omega^-$ \\
\hline
 $\sigma_{\text{Baryon}} (\text{MeV})$ &44(3)(3)	&31(1)(2)	&25(1)(1)	&15(1)(1)	
 &29(9)(3)	&18(6)(2)	&10(3)(2)	&5(1)(1)	\\
\hline 
\end{tabular}
\end{table}

Following the same procedure as for the mesons, we have carefully 
categorized the higher excited states for the baryons as well. For the 
nucleon and $\Delta$ resonances we have checked that their $\sigma$ terms, extracted from the lattice data~\cite{Edwards:2011jj} are within 1$\sigma$ 
with the corresponding ground state values. Excited strange baryons with strangeness content $S=1, 2, 3$ have been included by setting their 
corresponding $\sigma$ term same as the ground state baryons $\Sigma^*, 
\Xi^* \text{and}~\Omega^-$ respectively. We end this section by summarizing 
the $\sigma$ terms used in our calculations for the ground states and 
higher excitations in the various quantum number channels for baryons and 
mesons in Table~\ref{table:sigmasummary}.

\begin{table}[h]
\caption{\label{table:sigmasummary} A summary of the $\sigma$ terms for 
various meson and baryon channels used in this work.  }
\begin{center}
\begin{tabular}{|c|c|c|c|}
\hline
Mesons & Meson $\sigma$ term (MeV) & Baryons & Baryon $\sigma \text{-term (MeV})$\\
\hline
Pseudoscalar ($\pi(135), K(495)$) & NLO $\chi$PT~ 
Refs.~\cite{Bali:2016lvx,RBC-UKQCD:2008mhs} & Octet 
($N, \Lambda, \Sigma, \Xi$) & Table~\ref{table:sigma} \\
\hline
$\eta(547),~\eta^{'}(957),~\rho (770), K^*(892)$ 
& Refs.~\cite{Bali:2015gji, Guo:2016zos,Bali:2021qem} & 
Decuplet ($\Delta, \Sigma^*, \Xi^*, \Omega^-$) & Table~\ref{table:sigma} \\
\hline 
Resonances $|S|=0, |I|=1$ , ($\pi, \rho, a, b$) & Similar to $\rho(770)$  & 
Excited states of $N$ and $\Delta$ & Similar to $\sigma_{N}$ and $\sigma_{\Delta }$ \\
\hline
Resonances $|S| \neq 0$, ($K, K^*$) & Similar to $K^*$ & Excited states 
labeled as $\Lambda, \Sigma, \Sigma^*$ & 
Similar to $\Sigma^*$ \\ 
\hline
Resonances $|S|=0, |I|=0$, ($\phi, \eta, f$) & Similar to ground states 
($\eta,\eta', \omega, \phi$)  &  
Excited states labeled as $\Xi, \Xi^*, \Omega$ & Similar to $\Xi^*$ and $\Omega$ \\
\hline
\end{tabular}
\end{center}
\end{table}

\subsection{The condensate in the chiral limit}
\label{Sec:Condchirallim}
In QCD with two light quark flavors, the crossover transition goes over 
to a real phase transition in the chiral limit. This happens because the 
critical fluctuations in the free energy start dominating over the 
regular (analytic in the light quark mass) terms towards the chiral 
limit. Within a HRG model, the information about critical fluctuations 
may be entirely missing in the absence of additional inputs about the 
thermal widths, particularly of the pseudo-Goldstone partners $\pi, 
f_0$. Unlike in QCD, the $\langle \bar\psi\psi\rangle_T$ within the HRG 
model will not be zero at the critical temperature $T_c^0$.

In order to calculate the chiral condensate we need to calculate the 
derivatives of the square of the hadron masses with respect to the light 
quark mass, but now in the limit $m_l\to 0$. These derivatives for the 
pion and kaon mass squared can be calculated using Eq.~(\ref{Eq.GMOR}), 
which for the $m_l\to 0$ case read:
\begin{eqnarray}
\frac{\partial M_\pi^2}{\partial m_l}=2B~,~
\frac{\partial M_K^2}{\partial m_l}= 2B~B_K(m_s)m_s\frac{\lambda_1(m_s)+\lambda_2(m_s)}{F^2}~.
\label{Eq.GMORc}
\end{eqnarray}
From the latest FLAG review~\cite{Aoki:2021kgd}, the value of 
$B=\Sigma/F^2=2.76$ GeV, where $F$ and $\Sigma$ are the pion 
decay constant and light quark condensate respectively measured in the 
chiral limit of SU(2)$~\chi$PT. Since pions are the Goldstone modes of 
QCD with two massless flavors, they will have zero mass. The kaon and 
$\eta$ meson masses decrease only slightly in the chiral limit as their 
masses are determined primarily by the strange quark mass. We then 
calculate the chiral condensate defined in Eq.~(\ref{eq.defppbar}) in 
the limit $m_l\to 0$ and normalize by its zero temperature value.

Except for the pions and kaons the remaining hadrons and resonances 
have been included in the definition of renormalized chiral condensate 
by rearranging Eq.~(\ref{eq.sub_psibarpsi}) such that it could be 
generalized easily to the chiral limit. This is done through the 
relation
\begin{eqnarray}
\lim_{m_l\to 0}~{\sum }_{\alpha}
\frac{\partial P_\alpha}{\partial m_l}=\lim_{m_l\to 0}
\sum_{\alpha}\frac{\partial P_\alpha}{\partial M_\alpha}
\frac{\partial M_\alpha}{\partial M_\pi^2}
\frac{\partial M_\pi^2}{\partial m_l}
=\frac{2 B}{M_{\pi_{\text{phys}}}^2}~\sum_{\alpha}
\frac{\partial P_\alpha}{\partial M_\alpha}~\sigma_\alpha~.
\label{Eq.sub_B_chiral}
\end{eqnarray}
Here we have used the same definition of the $\sigma$ term as 
introduced for the baryons in Eq.~(\ref{Eq:SigmaDef}) and hence we had 
to normalize the right-hand side of Eq.~(\ref{Eq.sub_B_chiral}) by the 
square of the physical pion mass. The derivative term 
$\lim_{m_l\to 0}\partial M_\pi^2/\partial m_l=2B$ and the sum in Eq.~(\ref{Eq.sub_B_chiral}) is over all the hadron species except for 
the pions and kaons. In the Eq.~(\ref{Eq.sub_B_chiral}), we have used 
the masses of the ground states of the baryon octet and decuplet which 
were extracted in the chiral limit from Ref.~\cite{Copeland:2021qni}. 
For the $\rho$ meson, the fit to its mass data as a function of 
$M_\pi^2$, as shown in Appendix~\ref{appdx:a}, yielded $M_\rho=690(18)$ 
MeV in the chiral limit which is about $10\%$ lower than its physical 
mass. However, we could not find any data for the masses of most other 
resonances in the chiral limit, but instead took their physical masses. 
To estimate the systematic uncertainty due to this approximation, we 
varied the mass of these resonances by $10\%$ but found a $\approx 1\%$ 
change in the value of the temperature at which the chiral condensate 
vanishes. We thus anticipate that the effect of the mass modification 
of these resonances will be tiny in the chiral limit.

\subsection{Estimating the errors in our results}
The major source of error in the calculation of the renormalized chiral 
condensate comes from the errors in estimating the $\sigma$ terms. In 
this work, we have improved upon the errors considerably by estimating 
the $\sigma$ terms from the lattice QCD data for the $\rho, K^*$ and 
the $\eta,\eta^{'}$ mesons and the excited states of pions. For the 
baryon sector, the $\sigma$ terms for the octet and decuplet ground 
states have been extracted to a very good 
precision~\cite{Copeland:2021qni}. However, large uncertainty exists 
about the values of the $\sigma$ terms of the excited baryon states. 
We show in Appendix~\ref{appdx:a} how extraction of the $\sigma$ term 
of the excited nucleon state is prone to larger errors due to 
uncertainty in calculating the masses of these states. Although we 
have assumed that the higher mass resonances have a similar value of 
$\sigma$ term as their ground states hadrons, various studies hint 
that their $\sigma$ terms may deviate from those of their 
corresponding ground states \cite{Dudek:2010wm, Dudek:2013yja}. To 
reliably account for this uncertainty, we have taken the relative 
errors in the $\sigma$ terms of all the excited states of the baryons 
to be $50\%$. We have found that this variation does not result in a 
significant change in the renormalized chiral condensate. Due to their 
relatively heavier masses, the condensate is mainly dominated by the 
ground state hadrons in the temperature region till $155$ MeV. 
Although the resonances have a significant cumulative contribution 
to the condensate, the uncertainties in their $\sigma$ terms effect 
the estimates of $T_c$ by less than $1\%$.

The largest sources of error are still due to the uncertainties in the 
derivatives of the pion and kaon mass with respect to $m_l$, which we 
have estimated within the $\chi$PT.  For example, the relative error 
for the $m_l$ derivative of the pion mass is almost $6\%$, which arises 
mainly due to the uncertainty in the low energy constant $B$. For 
kaons, the error is even higher $\approx23\%$, when we use 
Eq.~(\ref{Eq.GMOR}). This is because the low energy constants in 
$SU(3)~\chi$PT are not very precisely determined, and the majority, 
$\approx18\%$, of the error arises from the uncertainty in
$\Sigma_0^{1/3}$ MeV and $F_0$ that appear explicitly in the $m_l$ 
derivative of kaon mass. The errors due to these two SU(3) $\chi$PT low 
energy constants also contribute to the error that comes from the term 
$\lambda_1(m_s)+\lambda_2(m_s)$, which accounts for the remaining $5\%$ 
contribution to the error on the $m_l$ derivative of the kaon mass 
squared. As the ground state pseudoscalar mesons dominantly contribute 
in the chiral condensate at $T\approx 156$~MeV, these errors also 
dominate the total error of the renormalized chiral condensate and hence 
the estimation of $T_c$.

\section{Results}

\subsection{Comparing the HRG model results for the renormalized chiral 
condensate with lattice QCD}

\begin{figure}[hbt!]
\includegraphics[width=8.6cm]{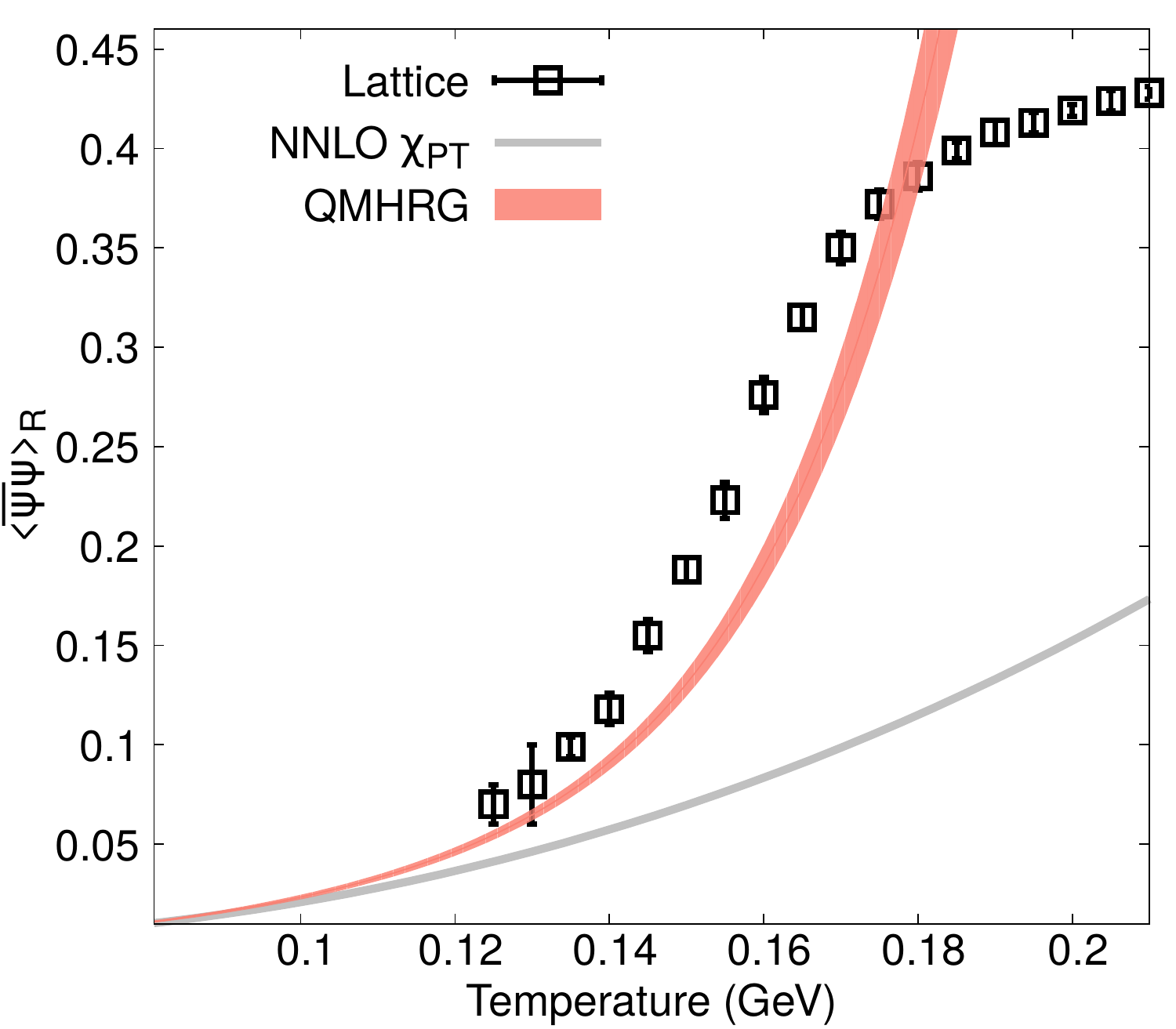} 
\includegraphics[width=8.6cm]{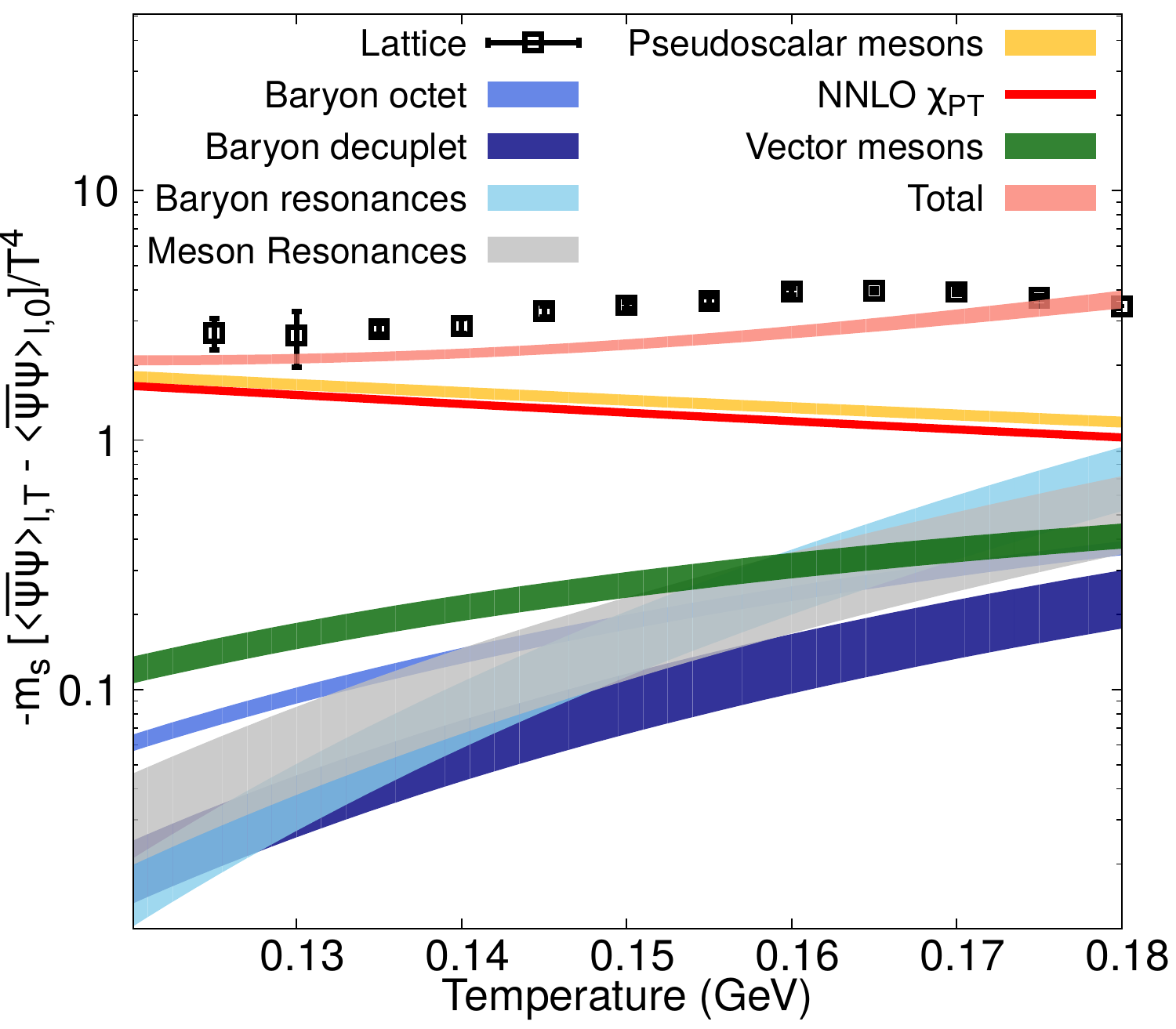} 
\caption{Left: Renormalized chiral condensate 
$\langle\bar{\psi}\psi\rangle_R=-\frac{m_l} 
{m_\pi^4}\left[\langle\bar{\psi}\psi\rangle_{l,T}-\langle\bar{\psi}
\psi\rangle_{l,0}\right]$ calculated within QMHRG is compared to the 
continuum extrapolated lattice data from Ref.~\cite{Borsanyi:2010bp}. 
Right: The relative contribution to the subtracted light quark 
condensate, $-m_s\left[\langle\bar{\psi}\psi\rangle_{l,T}-\langle\bar{\psi}
\psi\rangle_{l,0}\right]/T^4$, due to different meson and baryon channels 
is shown and the resultant total contribution within QMHRG (orange band) 
is compared to the lattice data from Ref.~\cite{Borsanyi:2010bp}.}
\label{Fg.qqbar_R}
\end{figure}

We compare our HRG model calculations with the continuum extrapolated 
lattice QCD results for $\langle\bar{\psi}\psi\rangle_{R}$, from 
Ref.~\cite{Borsanyi:2010bp}.  The lattice results were calculated for 
$2+1$ flavor QCD with staggered discretization for quarks with one-link 
stout improvement and physical quark masses such that $m_s/m_l=28.15$. 
To compare our HRG model calculation to the lattice data we 
have accordingly set the ratio of quark masses to the same value 
in the model as well.

We show the comparison of the renormalized chiral condensate as defined 
in Eq.~(\ref{Eq.psibarpsi_R}) calculated in lattice 
QCD~\cite{Borsanyi:2010bp} and the HRG model in Fig.~\ref{Fg.qqbar_R} 
(left). This quantity shows a smooth analytic rise as a function of 
temperature, as is evident from the continuum extrapolated lattice 
results from Ref.~\cite{Borsanyi:2010bp}, gradually saturating at 
$T\approx 200$ MeV. Since there is no real phase transition for 
physical quark masses, we do not observe any sharp jump at the chiral 
crossover transition temperature $T_c$. Our results calculated within 
the noninteracting HRG model agree well with the lattice results below 
$T=140$ MeV. Above this temperature, the HRG model results show a 
rapid rise with temperature, a trend similarly visible within the 
continuum extrapolated lattice data, however, the HRG estimates are 
lower than the lattice QCD results. 
As will be discussed later, while for $T>140$ MeV pion and kaons 
provide the dominant contribution to the renormalized chiral 
condensate, the relative contributions of the vector mesons and baryons 
start to increase near the chiral crossover, thus influencing the 
location of the corresponding pseudocritical temperature. The 
$\langle\bar{\psi}\psi\rangle_{R}$ calculated from HRG shows a 
diverging behavior at high temperatures $T\approx 170$ MeV, which is 
similar to the singularity in the Hagedorn 
spectrum~\cite{Hagedorn:1965st}, due to the density of states 
increasing exponentially with temperature. This is primarily driven due 
to a large number of excited states of different spin-parity channels 
in the baryon sector. There is no sign of saturation of 
$\langle\bar{\psi}\psi\rangle_{R}$ in the HRG model at high 
temperatures unlike on the lattice.

We have also compared our results with the calculations from NLO chiral 
perturbation theory in the two-loop approximation in the presence of 
a pion heat-bath taken from Refs.~\cite{Kaiser:1999mt,Holt:2014hma}. The 
$\chi$PT results agree with our HRG results at $T\lesssim 100$ MeV, 
beyond which the latter increases more strongly with temperature 
leading to a larger difference between the two.

A similar comparison of lattice QCD data with the HRG model predictions 
was shown in Ref.~\cite{Borsanyi:2010bp}. However, the uncertainty in the 
HRG predictions have been reduced by an order of magnitude in our results. 
This is possible now because of a more precise determination of $\sigma$ 
terms for the ground state baryons~\cite{Copeland:2021qni}. Furthermore, 
with a significant improvement in the calculation of $\rho$ meson mass
on the lattice as a function of the pion mass~\cite{Loffler:2021afv}, 
we could extract the corresponding $\sigma$ term more precisely.
Similar improvement of the $\sigma$ terms of the pseudoscalar isosinglet 
meson excited states has also led to a more accurate HRG model estimate 
of the renormalized chiral condensate.

To check the robustness of our predictions we have also calculated 
a different definition of the renormalized chiral condensate 
$-m_s\left[\langle\bar{\psi} \psi\rangle_{l,T}-\langle\bar{\psi}
\psi\rangle_{l,0}\right]/T^4$, shown in the right panel of Fig.~\ref{Fg.qqbar_R}. 
The temperature dependence of this quantity is similar to that of the trace 
anomaly calculated in Ref.~\cite{Bazavov:2014pvz}, and it is also useful in 
order to understand the relative importance of different hadron species 
contributing to the chiral condensate. 
As one can observe from Fig.~\ref{Fg.qqbar_R}, the heavier hadrons become 
important for $T>150$ MeV. These are the states which are responsible for 
the increase of this observable as a function of temperature, since the 
contribution from the pseudoscalar states monotonically decreases beyond 
$T\approx 100$ MeV. We can also see that the contributions of the 
higher-lying baryon and meson resonances are important for $T>150$ MeV
and become comparable in magnitude to the individual contributions due 
to octet and decuplet baryons and vector mesons. The quark mass dependence 
of these excited hadron resonances are not very well constrained by 
lattice QCD calculations and the errors on the corresponding $\sigma$ 
terms are comparatively large. Hence the error band for this observable 
also increases with temperature. Based on the discussion in Appendix A
we assign a generous $50\%$ relative error for the $\sigma$ terms of these 
higher-lying meson and baryon resonances. In Appendix B we further scrutinize 
the relative contribution of these resonances to the chiral observables.

\begin{figure}[hbt!]
\includegraphics[width=8cm]{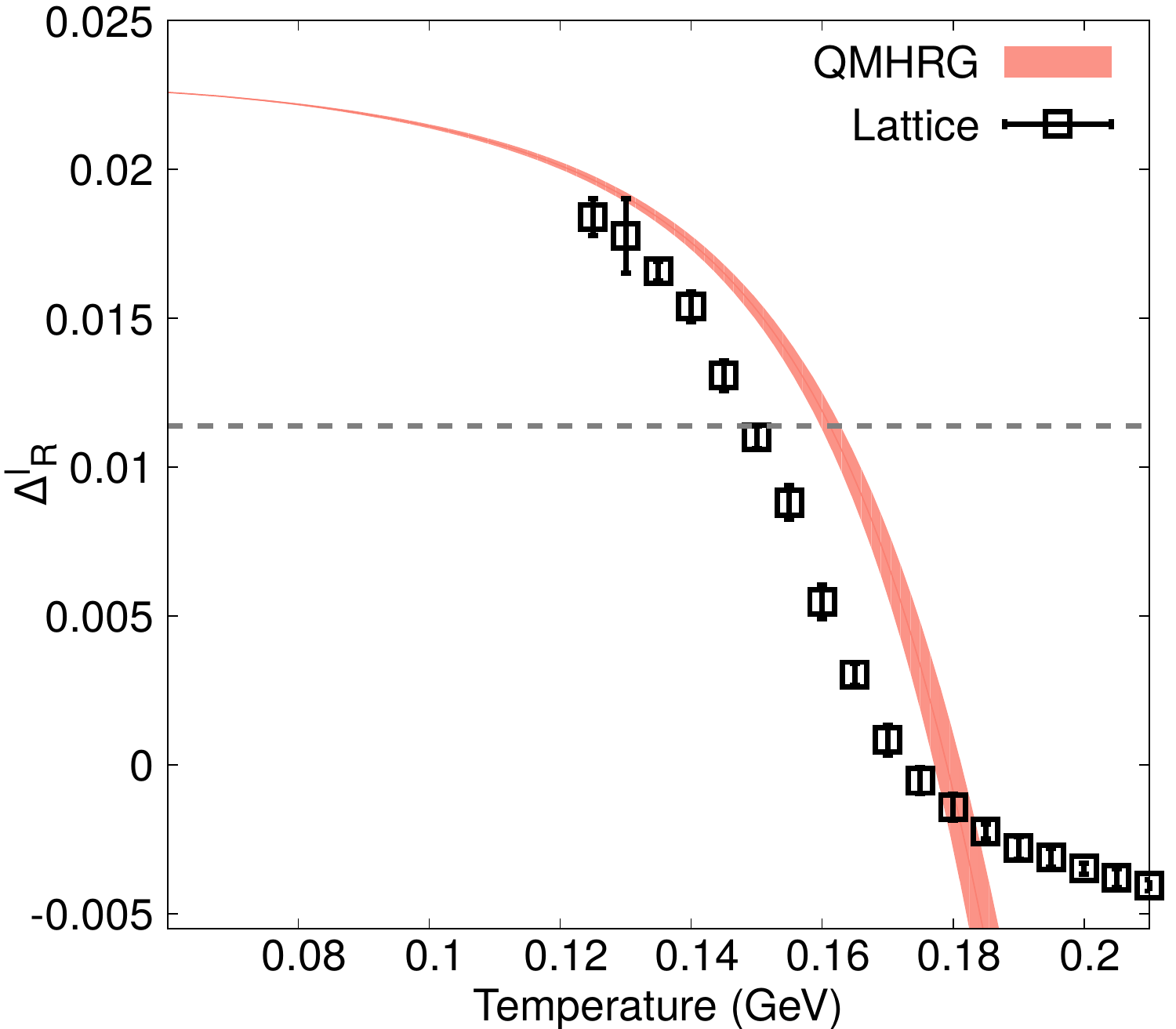} 
\includegraphics[width=8cm]{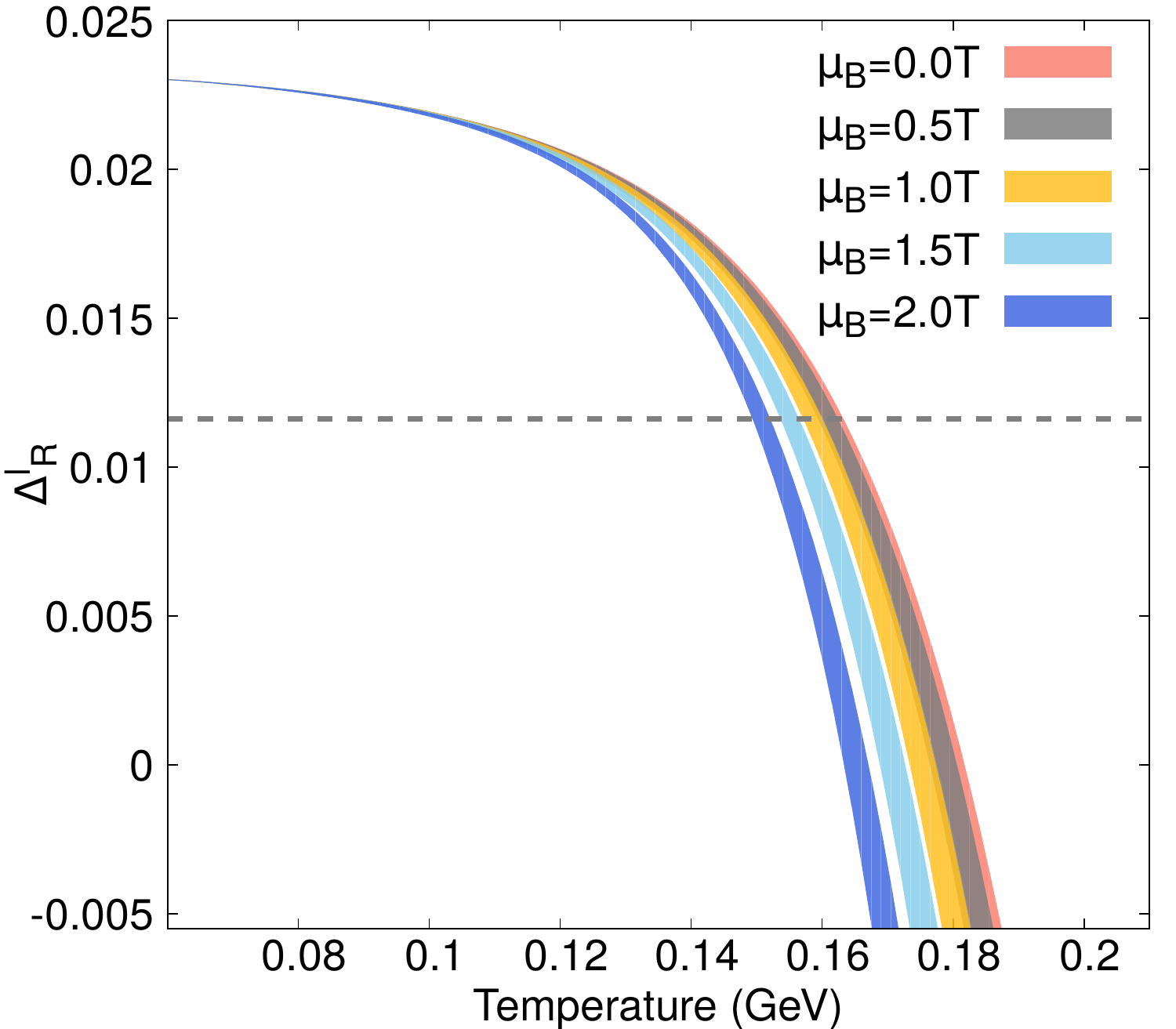} 
\caption{Left: $\Delta^l_R$, as defined in Ref.\cite{Bazavov:2011nk},
$\Delta^l_R=d+ m_sr_1^4 \left[\langle\bar{\psi}\psi\rangle_{l,T}-
\langle\bar{\psi}\psi\rangle_{l,0}\right]$ is compared to the lattice 
data taken from Ref.~\cite{Borsanyi:2010bp}. We have used $d=0.022791$ and 
$(r_1 m_\pi)^4=2.23\times 10^{-3}$ following \cite{Bazavov:2011nk}. 
The dotted line denotes the half value of $\Delta^l_R$ compared to its 
magnitude at the lowest temperatures, which is used as a criterion to 
determine $T_c$.
Right: $\Delta^l_R$ has been plotted for different values of baryon 
chemical potentials $\mu_B/T\leq 2$. In both the figures an estimate
of $T_c$ is obtained from where the magnitude of $\Delta_R^l$ falls to 
half its zero-temperature value.}
\label{Fg.deltalR}
\end{figure}

In the left panel of Fig.~\ref{Fg.deltalR}, we show the temperature 
dependence of another definition of the renormalized chiral condensate, 
$\Delta^l_R$ as calculated from the HRG model and compared with the 
lattice data.  This quantity is related to the above definition of the 
chiral condensate through $\Delta^l_R=d-(m_s/m_l)(r_1 m_\pi)^4
\langle\bar{\psi}\psi\rangle_R$. The study of $\Delta^l_R$ is motivated 
by our desire to extract the pseudocritical temperature corresponding 
to the chiral crossover transition, within our HRG model calculations. 
A comparison with the lattice QCD data for $\Delta^l_R$ leads to 
similar conclusions as earlier, namely for $T \le 140$ MeV the difference 
between the lattice and HRG model result is small, while for $140<T<160$ 
MeV the HRG model results for $\Delta^l_R$ drop slower than the lattice QCD 
results. At still higher temperatures the lattice results for $\Delta^l_R$
flattens off signaling a change in the degrees of freedom, while the HRG 
model data keep decreasing.

Let us now discuss the technique of determination of the pseudocritical 
temperature. In lattice calculations, the pseudocritical temperature is 
defined as an inflection point of the renormalized chiral condensate or 
the peak of the chiral susceptibility as a function of temperature. Due 
to the breaking of the exact chiral symmetry due to a finite quark mass, 
the value of the pseudo-critical temperature will depend on the quantity 
used to define it, including the normalization of the corresponding 
quantity~\cite{Lahiri:2021lrk}. In a recent lattice QCD study with 
highly improved staggered quark discretization, it was found that the 
spread of the pseudocritical temperature obtained from chiral 
observables is surprisingly small, $\approx1.5$ MeV. A somewhat larger 
spread was obtained in another study with stout-improved staggered 
fermions~\cite{Borsanyi:2010bp}. Here the spread was about $8$ MeV when 
results from the chiral susceptibility were taken into account. The main 
difference is due to a different normalization of the chiral 
susceptibility suggested in Ref.~\cite{Borsanyi:2010bp}.

The HRG model results on the renormalized chiral susceptibility, on the 
other hand, do not have an inflection point. This is obvious from 
Fig.~\ref{Fg.deltalR}. Therefore, one cannot define a pseudocritical 
temperature from the HRG model results alone. However, as mentioned in 
the Introduction, the breaking of the chiral symmetry plays a crucial 
role in hadron physics. If at some temperature the renormalized chiral 
condensate becomes small compared to its vacuum value the hadronic 
matter should gradually melt to a new state. We do not know a $priori$ 
how small the chiral condensate should be for this to happen because of 
the inherent nonperturbative nature. Lattice QCD calculations show 
that around the pseudocritical temperature the value of the chiral 
condensate $\Delta_l^R$ is around half of its vacuum 
value~\cite{Bazavov:2011nk,Bazavov:2013yv}. Motivated from this insight 
from QCD, we therefore set the same criterion to estimate $T_c$ within 
the HRG model calculations. With this procedure we estimate the 
pseudocritical temperature $T_c=161.2\pm 1.7$ MeV,  which is only 
$2 \sigma$ away from the recent high precision lattice QCD results on 
$T_c$ in the continuum limit~\cite{HotQCD:2018pds,Borsanyi:2020fev}. 
This is not completely surprising as the temperature dependence of the 
renormalized chiral condensate in the HRG model and in lattice QCD is 
similar, and the obtained estimate of $T_c$ is not completely 
independent of the lattice determination. However, this analysis 
represents a marked improvement over earlier estimates $T_c$ obtained 
from the HRG model~\cite{Toublan:2004ks,Tawfik:2005qh,Jankowski:2012ms} 
or from $\chi$PT~\cite{GERBER1989387,GarciaMartin:2006jj}, where the 
central values are more than $10 \sigma$ away from the latest lattice 
result. 

It is more interesting to extend the above analysis to nonzero baryon 
chemical potential and see how the transition temperature defined 
within the HRG model depends on baryon density. For that purpose, we 
calculate the temperature dependence of $\Delta^l_R$ for different 
values the baryon chemical potential $0\leq{\mu_B}/{T}\leq 2$. The 
corresponding results are shown in Fig.~\ref{Fg.deltalR} (right). At 
finite netbaryon densities the contribution from baryon states and 
resonances is expected to become more decisive,  which results in the 
$\Delta^l_R$ decreasing faster with increasing temperature compared to 
the $\mu_B=0$ case, as can be seen from Fig.~\ref{Fg.deltalR} (right). 
In turn, this leads to a decrease in the pseudocritical temperature 
with increasing baryon density as expected. However, with the increase 
in netbaryon density, repulsive interactions between baryons will be 
more probable and hence this simple noninteracting HRG model will 
cease to effectively describe the QCD medium. The effect of repulsive 
interactions on the pseudocritical temperature will be discussed in 
detail in our forthcoming publication~\cite{dps}.

\subsection{Curvature of the chiral crossover line from the HRG model}

As the next step we will investigate the dependence of the chiral 
crossover line as function of the baryon chemical potential $\mu_B$ 
within the HRG framework. As discussed in the previous sub-section we 
expect that the chiral crossover temperature will decrease with 
increasing $\mu_B$. We define the pseudocritical temperature at 
nonzero $\mu_B$ again, as the temperature where the renormalized chiral 
condensate drops by factor of 2, since we do not expect qualitative 
change in the chiral crossover unless $\mu_B$ is too large. At 
moderately large values of $\mu_B$ the pseudocritical temperature can 
be written as
\begin{eqnarray}
\frac{T_c (\mu_B)}{T_c(0)}=1-\kappa_2 \left(\frac{\mu_B}{T_c(0)}\right)^2-
\kappa_4 \left(\frac{\mu_B}{T_c(0)}\right)^4 + \mathcal{O}(\mu_B ^6)~.
\label{Eq.curvature}
\end{eqnarray}
The lattice calculations of the curvature $\kappa_2$ and its higher 
order corrections $\kappa_4$ are quite 
challenging~\cite{Steinbrecher:2018jbv}. For several years $\kappa_2$ 
measured using different lattice discretizations and systematics, did 
not agree with each other~\cite{Kaczmarek:2011zz, Bonati:2014rfa}, which 
was recently understood to be due to the fact that the continuum limit 
needs to be taken carefully~\cite{Bonati:2018nut}. Presently all 
continuum extrapolated lattice measurements of $\kappa_2$ from 
different groups using Taylor expansion~\cite{HotQCD:2018pds} or from 
imaginary chemical potential techniques~\cite{Bonati:2018nut, 
Borsanyi:2020fev} agree quite well with each other. However, the 
extraction of $\kappa_4$ remains a challenge since it involves delicate 
cancellation between noisy operators, due to which the signal-to-noise 
ratio in this observable is extremely low~\cite{HotQCD:2018pds, 
Borsanyi:2020fev}.

\begin{figure}[hbt!]
\includegraphics[width=8cm]{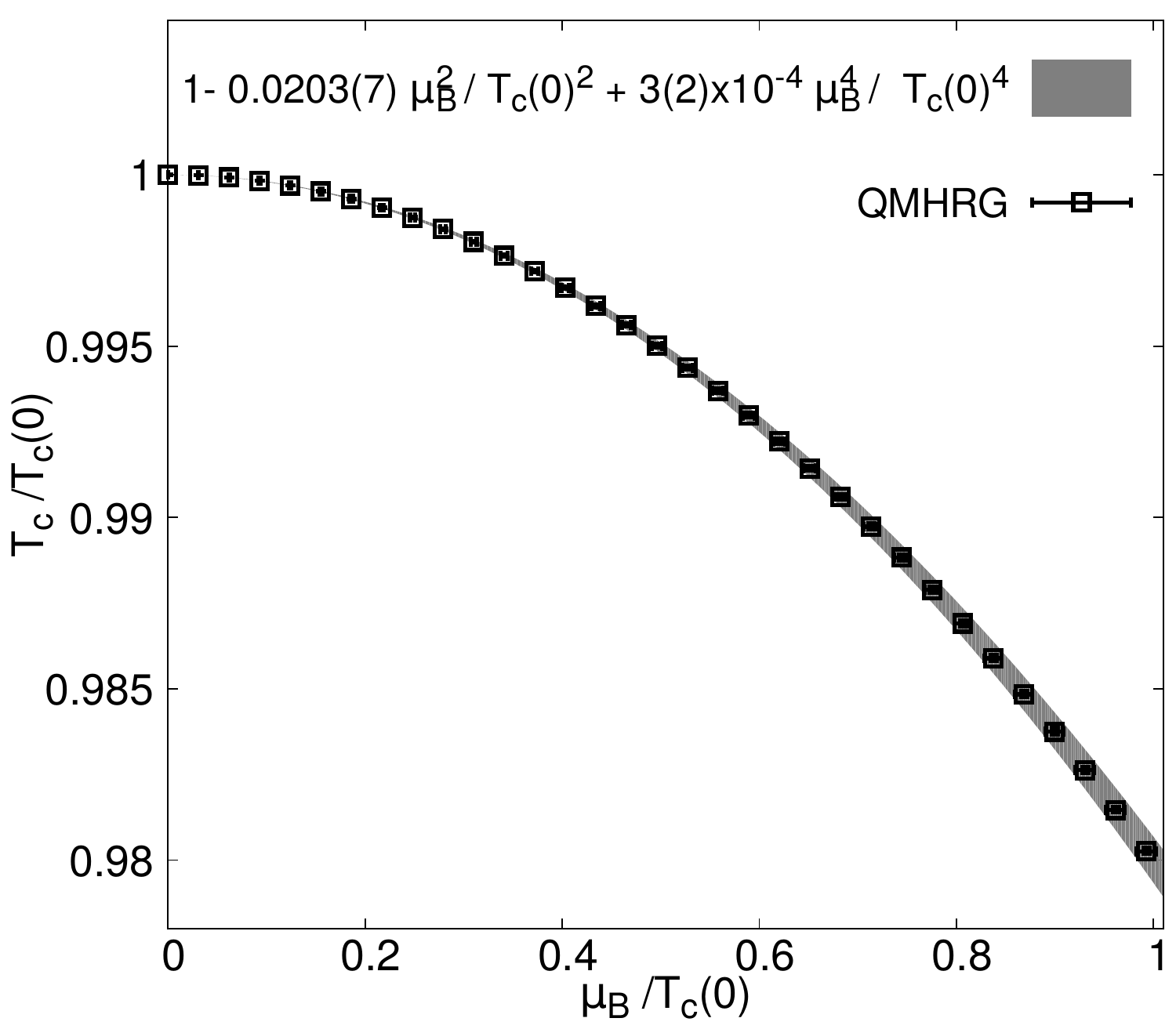}
\includegraphics[width=8cm]{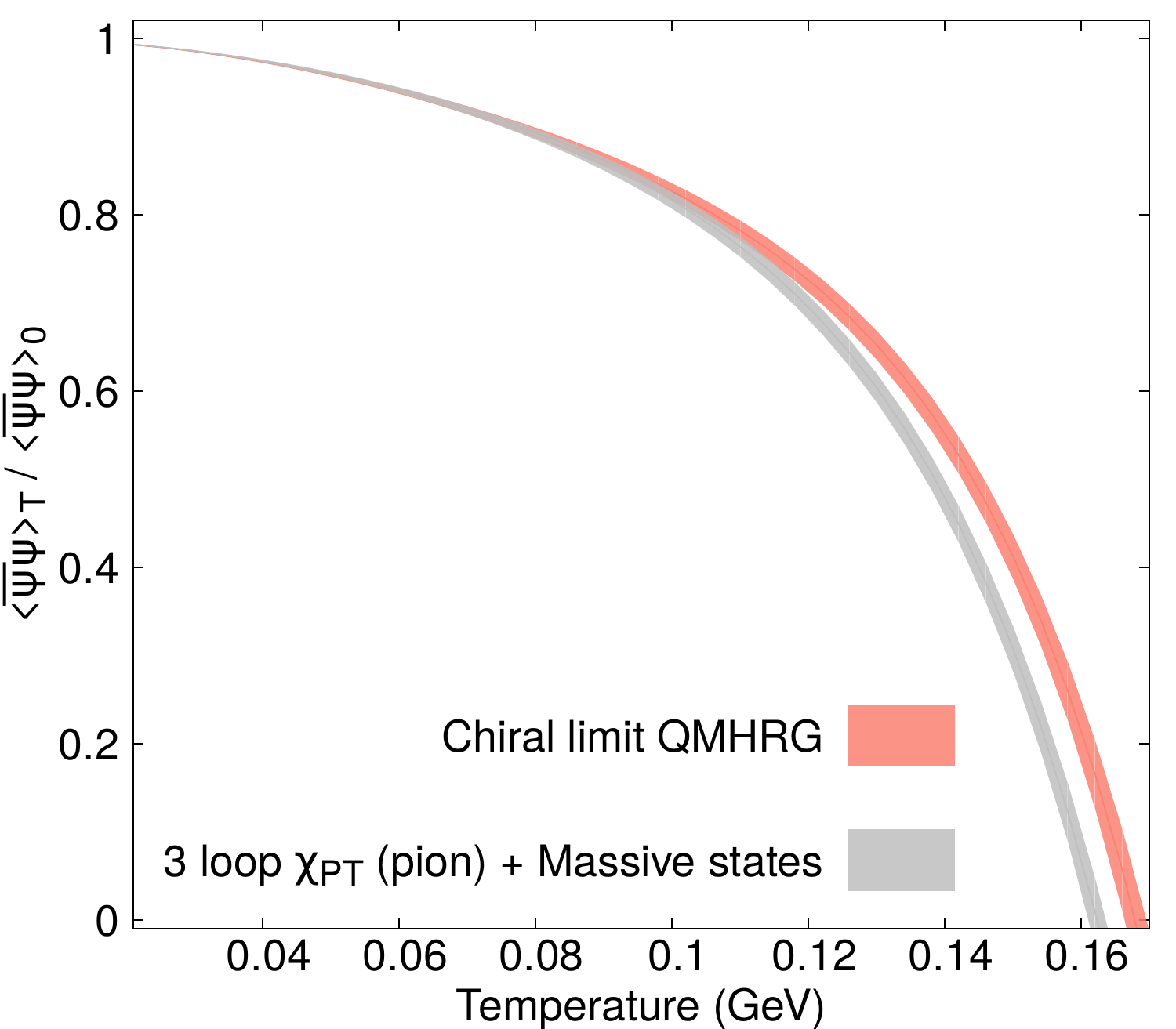}
\caption{Left: A fit to the pseudocritical line between $\mu_B/T 
\lesssim 1$ gave $\kappa_2=0.0203(7)$ and  $\kappa_4=-3(2)\times 10^{-4}$. 
Right: The ratio of the chiral condensate at a finite temperature to its 
zero temperature value, compared between our calculation within QMHRG 
(orange band) and three-loop $\chi$PT results up to $\mathcal{O}(T^6)$ 
from Ref.~\cite{GERBER1989387} which are augmented by the contribution 
of heavier hadrons with the spectrum that has been used in our work 
(shown as the gray band).}
\label{Fg.T-muB}
\end{figure}

In this context, we will study how well we can calculate the 
curvature terms $\kappa_2$ and $\kappa_4$ within the HRG model. For 
that, we first extract the pseudocritical temperature as a function of 
the baryon chemical potential $\mu_B$. We have then fitted the 
$T$-$\mu_B$ crossover curve with the ansatz in~Eq.~(\ref{Eq.curvature}) 
for values of $\mu_B /T < 1$. The resultant fits yielded a 
$\kappa_2=0.0203 \pm 0.0007$ within the HRG model which also is in a 
good agreement with the $\kappa_2=0.0150(35)$ extracted from the 
continuum extrapolated (subtracted) chiral condensate calculated in 
lattice QCD~\cite{Steinbrecher:2018jbv}. However, our extracted value 
of $\kappa_4=-3( 2) \times 10^{-4}$ is again quite noisy, consistent 
with the findings from lattice QCD~\cite{HotQCD:2018pds, 
Borsanyi:2020fev}. At small baryon densities, the curvature of the 
pseudocritical line is mainly determined by $\kappa_2$, which makes the 
extraction of $\kappa_4$ so difficult. To extract higher-order curvature 
coefficients one needs to extend our analysis to $\mu_B/T>2$, where it 
is anticipated that this simple noninteracting HRG approximation may 
break down due to repulsive baryon interactions. We summarize our main 
findings on the curvature coefficients in the left panel of 
Fig.~\ref{Fg.T-muB}. 

\subsection{Temperature dependence of the chiral condensate in the 
massless limit }

The chiral transition in the limit when $m_l\to0$ becomes a 
second-order phase transition with $O(4)$ or a $U(2)_L\times U(2)_R$ 
critical behavior depending on whether or not the anomalous $U_A(1)$ 
part of the chiral symmetry remains broken 
substantially~\cite{Pisarski:1983ms, Butti:2003nu, Pelissetto:2013hqa, 
Grahl:2013pba, Sato:2014axa, Nakayama:2014sba, Sharma:2018syt}. A high 
precision lattice QCD calculation of the chiral phase transition 
temperature with highly improved staggered quarks gives a 
$T_c^0=132^{+3}_{-6}$ MeV~\cite{HotQCD:2019xnw}. More recently there is 
a lattice QCD calculation using Wilson fermions~\cite{Kotov:2021rah} 
reporting a $T_c^0=134^{+6}_{-4}$ MeV which in good agreement with the 
earlier estimate. The $T_c^0$ estimated in $2+1$ flavor QCD within the 
functional renormalization group method yields a slightly higher value 
of $T_c^0\approx 142$ MeV~\cite{Braun:2020ada} but is reasonably close 
to the lattice calculation.  Even though pions will contribute 
dominantly to the condensate in the chiral limit,  extracting the  
$T_c^0$ is not possible within the conventional QMHRG models which 
do not have the information of the precise critical universality
class~\cite{Pelaez:2015qba,GomezNicola:2020yhm}. It was recently 
shown within linear sigma models with $O(4)$ universality or in 
$U(3)~\chi$PT, that it is necessary to include finite temperature 
corrections to the $f_0(500)$ spectral function to accurately account 
for the temperature dependence of the scalar susceptibility near the 
transition~\cite{Ferreres-Sole:2018djq}.

The earliest effort towards extracting the chiral transition 
temperature from $\langle \bar \psi\psi\rangle_T=0$ within a purely 
hadronic model, the three-loop $\chi$PT at finite temperature, gave a 
$T_c^0\approx 190$ MeV~\cite{GERBER1989387}. This large disagreement 
with the latest lattice estimates of $T_c^0$ can be understood from 
the fact that the chiral condensate due to an interacting pion gas 
calculated at three loops i.e., $\mathcal{O}(T^6)$, is reduced to half by 
$T=150$ MeV~\cite{GERBER1989387} thus limiting the validity of the 
calculations beyond this temperature. Nonetheless the authors in 
Ref.~\cite{GERBER1989387} extended their calculations to higher 
temperatures, also including the effects of a dilute gas of heavier 
hadrons that the pions increasingly encounter. The presence of 
heavier hadrons  reduces the chiral condensate to zero at a lower 
temperature, $\approx170$ MeV, compared to a pure self-interacting pion 
gas. Since in this work we have a substantially improved the procedure 
for inclusion of heavier hadrons and resonances, we included the effect 
of these states on the chiral condensate in addition to the three-loop 
pion gas contribution. The corresponding ratio of chiral condensates 
$\langle \bar \psi\psi\rangle_T/\langle \bar \psi\psi\rangle_0$ 
is shown as a gray band in Fig.~\ref{Fg.T-muB}. We observe that the 
chiral condensate reduces to zero at a temperature $~162$ MeV, which is 
about $8$ MeV lower compared to the earlier observed $T_c^0$ in 
Ref.~\cite{GERBER1989387}. If we extend our earlier calculation of the 
chiral condensate within QMHRG model but now in the chiral limit, 
following the procedure outlined in Sec.~\ref{Sec:Condchirallim}, we 
observe it to fall to zero at a comparatively higher temperature 
$\approx168$ MeV. This can be understood from the fact that only the 
attractive resonant interactions between pions are included in our QMHRG 
model. With this analysis, we conclude that extending the existing 
$\chi$PT calculation beyond three loops and carefully including the 
contributions of heavier hadrons may reduce the estimated $T_c^0$, but 
this estimate will be still far from the lattice QCD result.

\section{Summary and outlook}

Motivated from several pieces of evidence of a reasonably good 
representation of QCD thermodynamics at temperatures $T\lesssim 0.9~T_c$ 
by a noninteracting gas of hadrons and resonances, we revisit its 
validity for chiral observables in this study. We provide an updated 
analysis of the renormalized chiral condensate mainly using extensive 
lattice data available for the masses of several hadrons and resonances 
as a function of the pion mass, and some recent updates of the 
so-called $\sigma$ terms for ground state baryons. Furthermore, we 
include the effects of additional resonances, mainly strange baryon 
resonances, which are not yet measured experimentally but predicted 
from quark model and lattice studies. With the systematic inclusion of 
these effects we observe that a noninteracting HRG description for the 
renormalized chiral condensate is adequate to describe QCD data from 
first-principles lattice studies, up to $T\lesssim 140$ MeV. However, 
using the fact that the lattice studies observe a drop in the value of 
renormalized chiral condensate to half its zero temperature value at 
the chiral crossover transition, we  made an estimate of the 
pseudocritical temperature $T_c$ within this simple noninteracting HRG 
model. The extracted $T_c=161.2\pm 1.7$ MeV is in good agreement with 
the latest continuum extrapolated lattice QCD result, although this 
determination requires insights from lattice studies.

Encouraged by this effort we further estimate the curvature of the 
critical line at vanishingly small baryon density, yielding 
$\kappa_2=0.0203(7)$, which is in excellent agreement with the lattice 
results, which is highly nontrivial. Furthermore, with the more 
precise knowledge of the hadron $\sigma$ terms, we could achieve about 
$8$ MeV reduction in the earlier estimation of the chiral transition 
temperature in a system comprising a self-interacting pion gas at 
three loops and a gas of noninteracting heavier 
hadrons~\cite{GERBER1989387}. However, from our study we conclude that 
extracting the chiral transition temperature in this model would 
require further inclusion of higher order interactions between pions. 
Nonetheless our study has a deeper implication, since it allows us to 
carry forward our calculations within the HRG model with physical 
masses at higher baryon densities $\mu_B/T>3.5$, which is currently 
out of bounds of the lattice calculations, and calculations are in 
progress in this direction~\cite{dps}.
 
\section*{Acknowledgments}
P.P. is supported by the U.S. Department of Energy, Office of Science, 
through Contract No. DE-SC0012704.  S.S. gratefully acknowledges 
support from the Department of Science and Technology, Government of 
India through a Ramanujan Fellowship. D.B. acknowledges fruitful discussions 
with Jishnu Goswami and Sumit Shaw.

\begin{appendix}

\section{SIGMA TERMS FOR MESONS AND BARYONS}
\label{appdx:a}
As discussed in the main paper, the $m_l$ derivative of the masses of
the ground state pions and kaons, could be very well  described within 
the $\chi$PT framework. However, the $\sigma$ terms of other mesons are 
not known very precisely. This is true for the $\sigma$ terms for some 
ground state mesons like the vector mesons and $\eta, \eta^{'}$ mesons. 
The $\sigma$ terms for excited baryon states are also not very well 
known. This fact contributes to the uncertainty in the estimation of 
the chiral condensate. In this section we will evaluate the $\sigma$ 
terms with as much precision as possible using the most recent values 
of the hadron masses available from extensive lattice QCD studies. We 
have collected the most precise values of the hadron masses as available 
in the literature, as a function of the pion mass along the lines of 
constant and physical strange quark mass. We then fit these data to the 
ansatz $M_{\alpha}(M_\pi)=m^{\text{exp}}_{\alpha}+b_{\alpha} 
\cdot M_\pi^2$ and extract its slope $b_{\alpha}$ corresponding to each 
hadron species labeled by $\alpha$. Since we need the $\sigma$ term at 
the physical point, we have taken the product of the slope $b_{\alpha}$ 
extracted after performing the fit with the square of the average 
physical pion mass $M^{\text{phys}}_{\pi}=138$ MeV. 

\begin{figure}[h]
\includegraphics[width=0.32\textwidth]{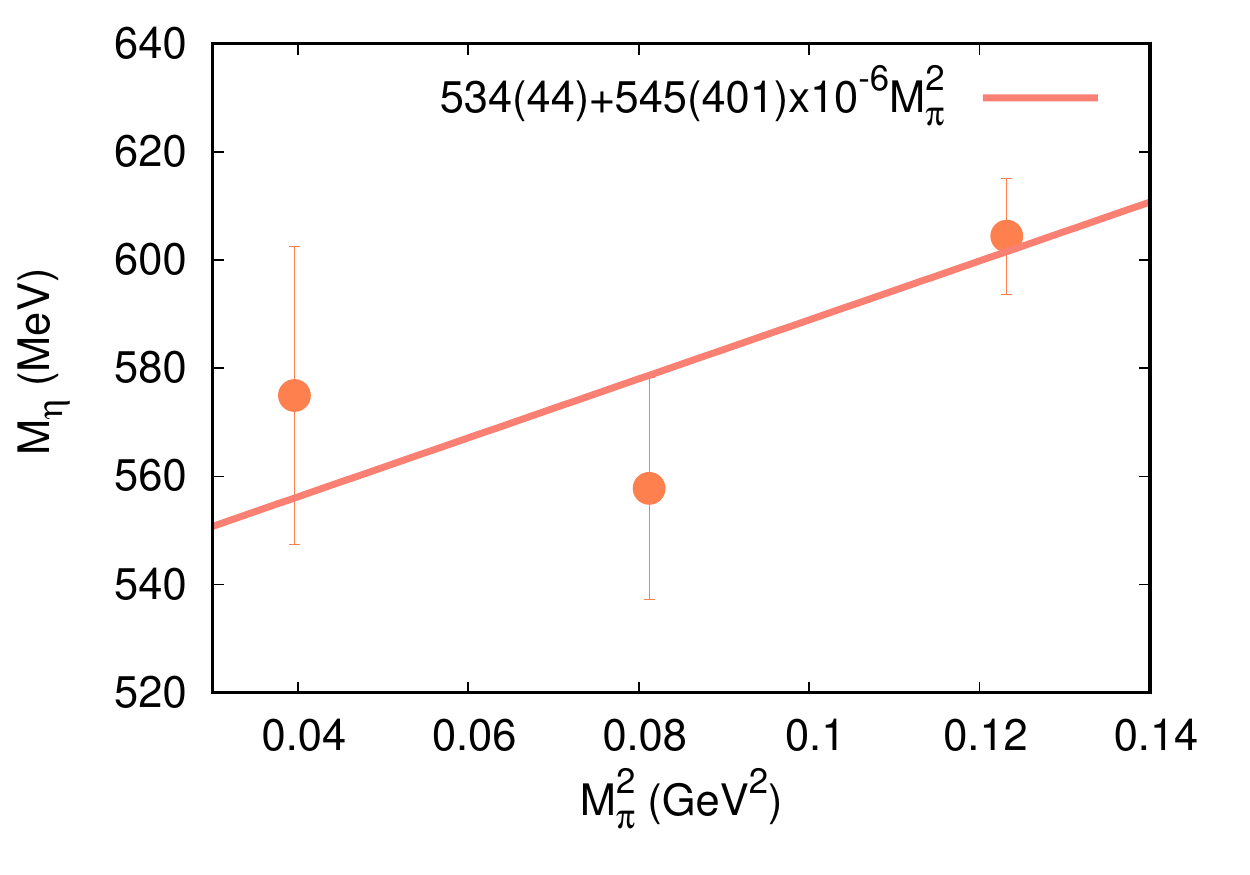}
\includegraphics[width=0.32\textwidth]{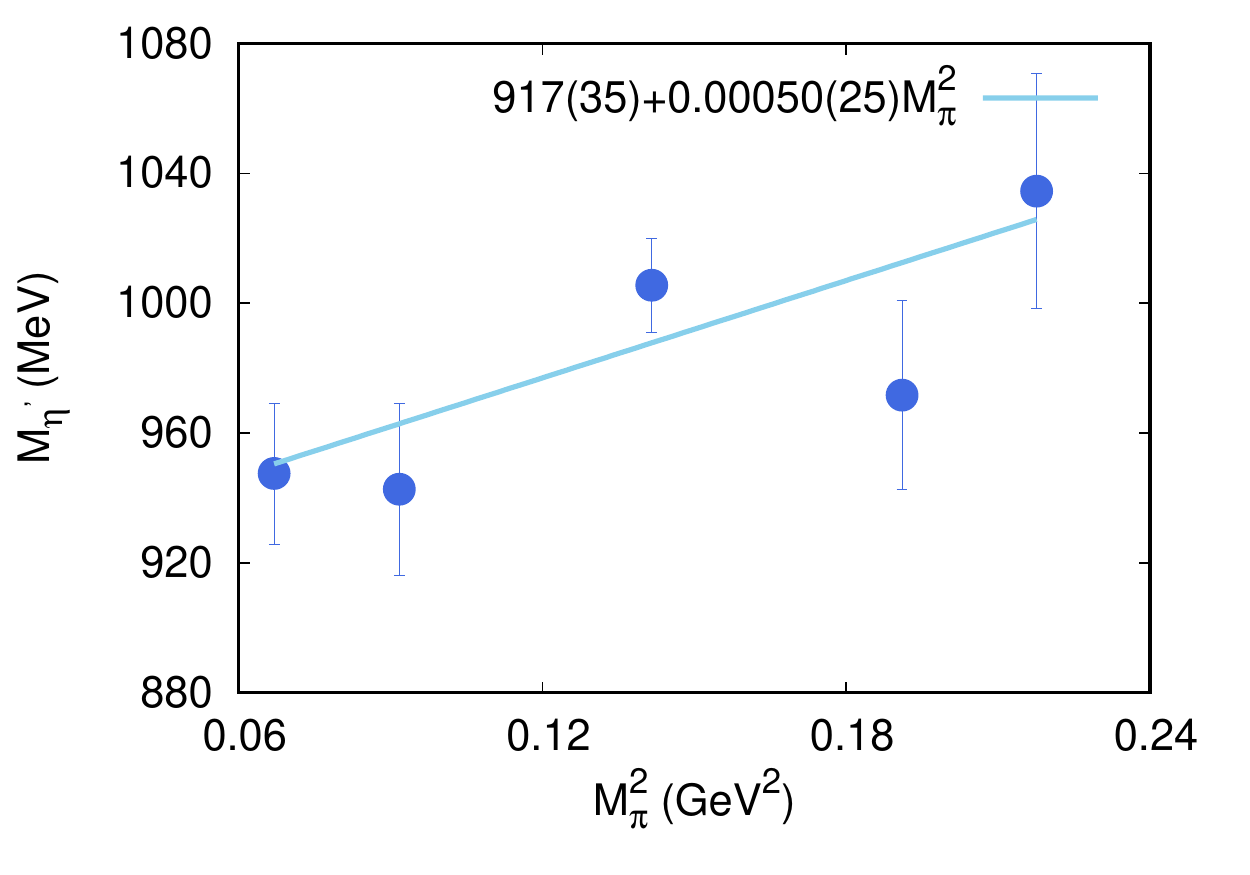} 
\includegraphics[width=0.32\textwidth]{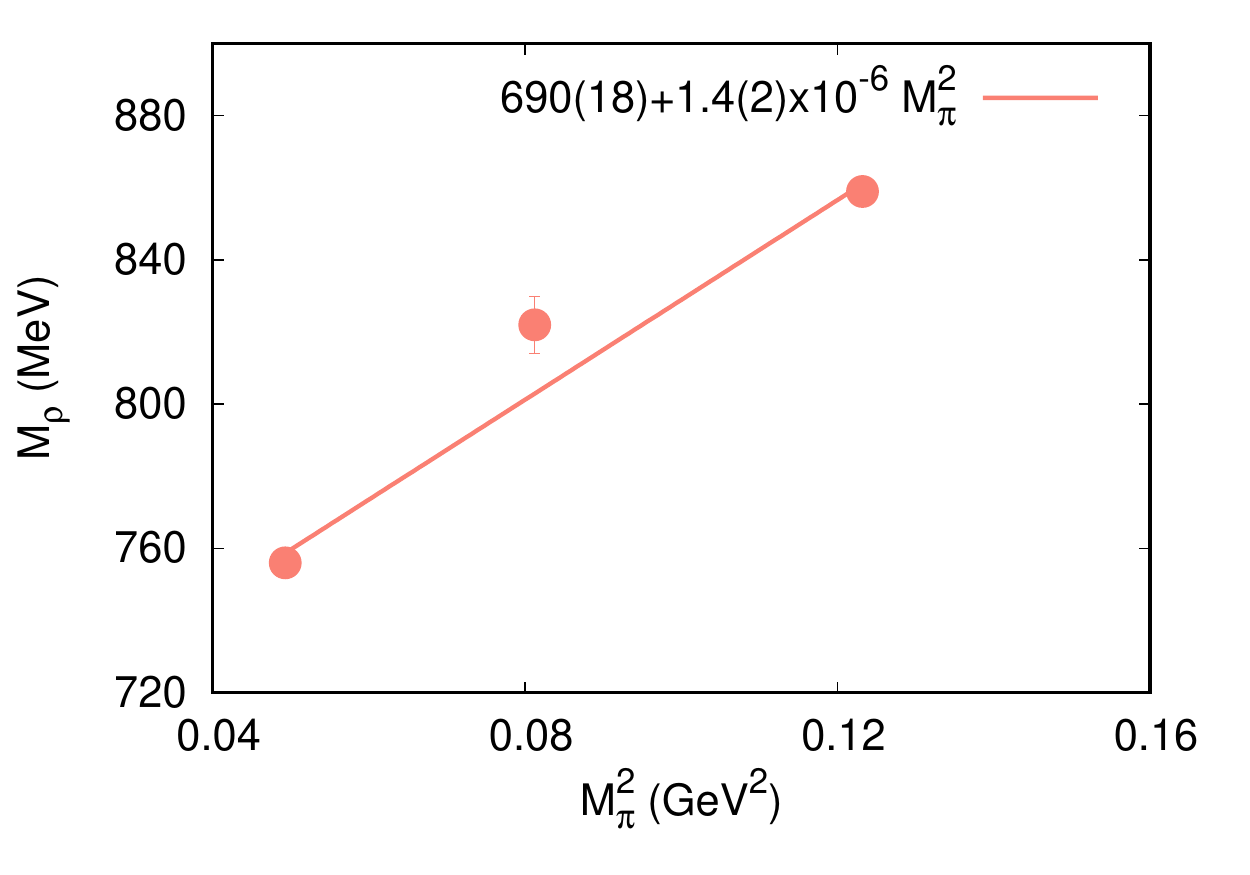} 
\caption{The left panel shows the fit to the mass of $\eta$ meson 
as a function of the square of the pion mass, with data taken from 
Ref.~\cite{Bali:2016lvx}. The middle and the right panels show the 
analogous fits for the $\eta^{'}, \rho$ mesons with the data taken 
from Refs.~\cite{Ottnad:2017bjt} and~\cite{Loffler:2021afv} respectively.
All fit results are in units of MeV.}
\label{fig:etaetaprho}
\end{figure}

\begin{figure}[h!]

\includegraphics[width=0.32\textwidth]{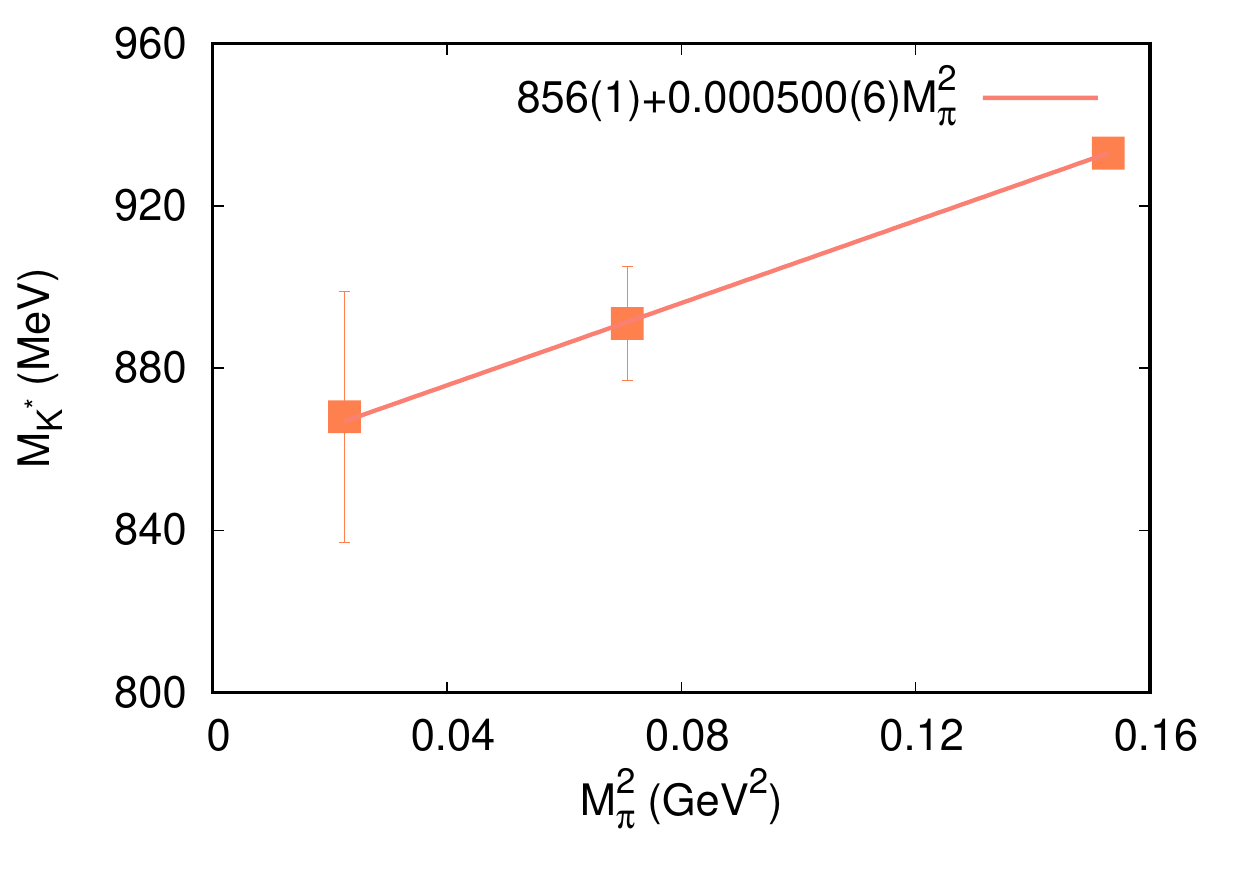} 
\includegraphics[width=0.32\textwidth]{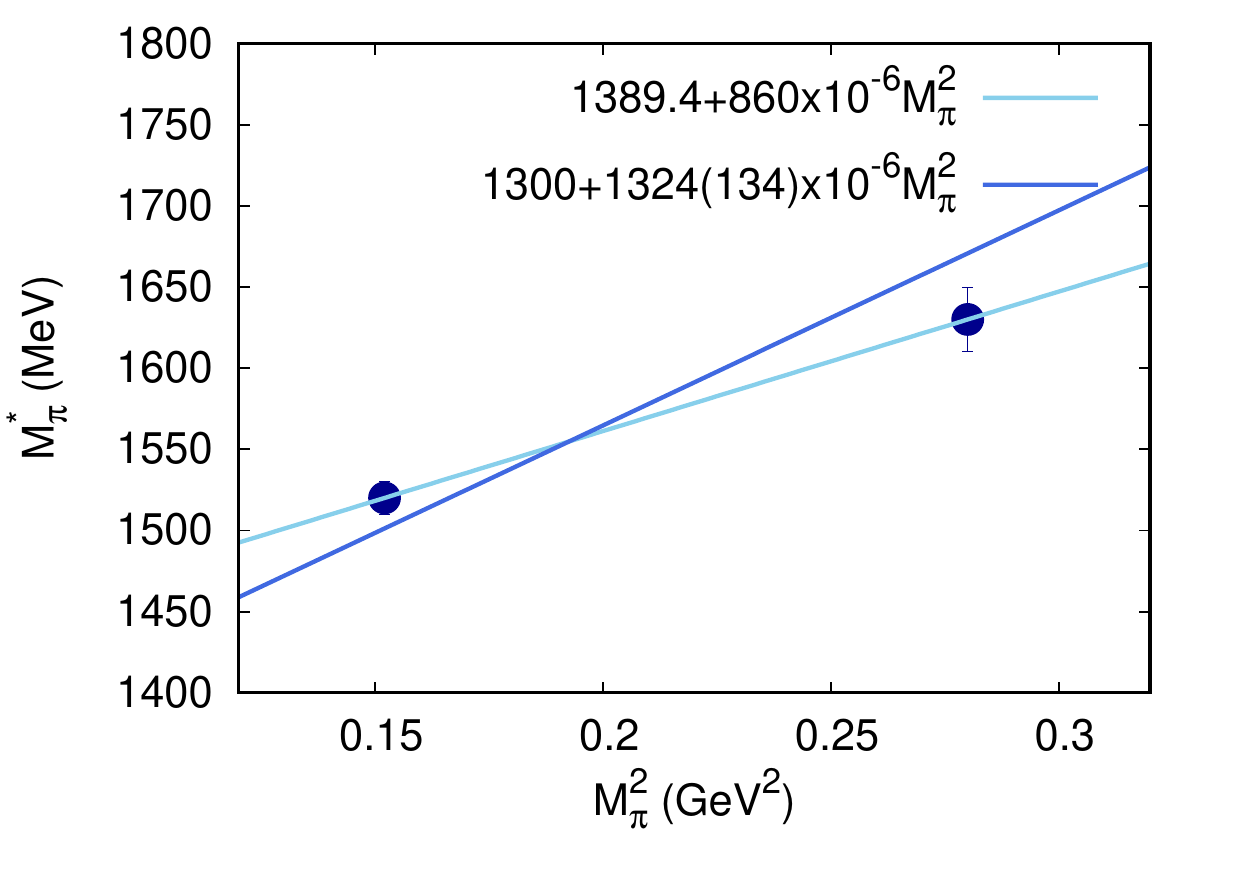} 
\includegraphics[width=0.32\textwidth]{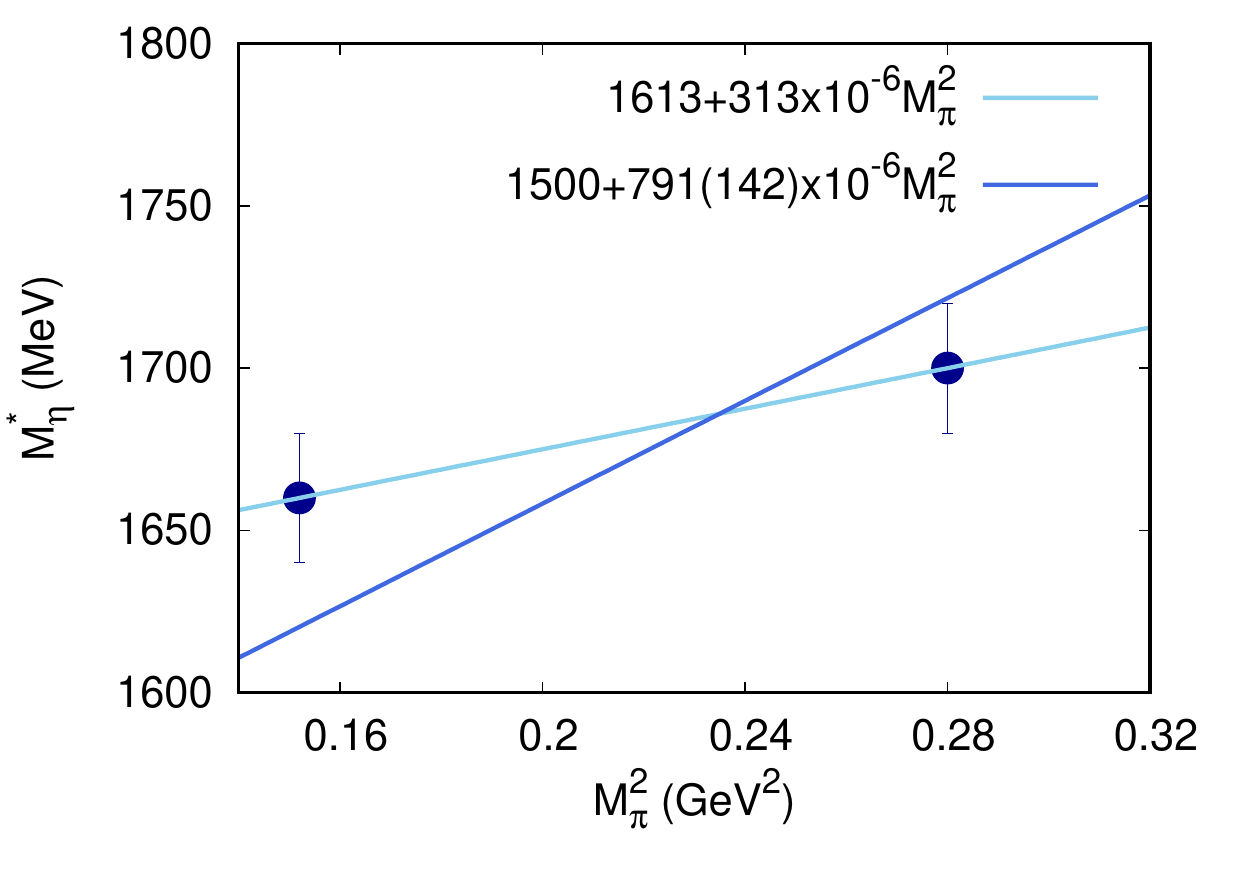} 
\caption{The fits to the mass of $K^*$-meson (left) 
for $2+1$ flavor QCD respectively shown as a function
of square of the pion mass. The data has been taken 
from Ref.~\cite{Bali:2015gji}. The middle and the 
right panel shows the fit to the mass of the excited 
states of the isoscalar $\pi$ and $\eta$-mesons as a 
function of the square of the pion mass, using the data from 
Ref.~\cite{Dudek:2013yja}. All fit results are in units of MeV.}
\label{fig:mesonresonances} 
\end{figure}

We first compile the dependence of the mass of the ground states 
of the isoscalar $J^{PC}=0^{-+}$ and vector $J^{PC}=1^{--}$ mesons, 
on the pion mass. The $\eta$ meson mass data were taken from 
Ref.~\cite{Bali:2016lvx}, and the results for $\eta^{'}$ and $\rho$ 
meson masses were compiled from Refs.~\cite{Ottnad:2017bjt, 
Loffler:2021afv}, respectively. From the fit shown in 
Fig.~\ref{fig:etaetaprho}, we extract a $\sigma=10\pm 7$ MeV for the 
$\eta$-meson, and a similar value of the $\sigma$ term for the 
$\eta^{'}$ meson i.e., $\sigma=9.5\pm 4.5~\text{ MeV}$.
For the $\rho$ mesons we could extract $\sigma=27\pm 4$ MeV with 
very good precision, which has enabled us to better constrain the 
$T_c$ from the HRG model. Using the data from Ref.~\cite{Bali:2015gji},
the fit of the mass of $K^*$ as a function of $M_\pi^2$, which is
shown in the left panel of Fig. \ref{fig:mesonresonances}.
From the fit, we extract a $\sigma_{K^*}=10\pm 1~\text{ MeV}$.
We also determined the $\sigma$ terms for the $\omega$ and $\phi$ mesons 
using the lattice data from Ref.~\cite{Dudek:2013yja}, where the mass 
spectrum was calculated in $2+1$ flavor QCD for two pion mass values 
$M_\pi=391,~524$ MeV respectively again along the lines of constant 
$m_s$ set to its physical value. We used the same procedure as above. 
The results are shown in Table.~\ref{table:octetdecupsigma}.

We next extract the $\sigma$ terms of the excited meson states. For the 
isotriplet excited states, we could again use the lattice data from 
Ref.~\cite{Dudek:2013yja}. Since there are only two data points in the 
fit, in order to estimate the goodness of the fit we have taken the 
experimentally determined masses as the intercept. This is a reasonable 
guess since we expect that the reduction of the mass of these excited 
states in the chiral limit will be very small, within $\approx 5$-$10\%$ 
of its physical value and not as dramatic as the pseudo-Goldstone pion 
modes. We also performed the fit without constraining the intercept 
in order to estimate the systematic uncertainty in the results of the 
$\sigma$ term. The results of the fits are shown in the middle panel of Fig.~\ref{fig:mesonresonances}. We find that the $\sigma$ terms 
corresponding to the first excited state of the pion are 
$\sigma=26(3),~16~\text{MeV}$ for the one- and two-term fits 
respectively. We see that the different estimates agree with each other 
if we assume the error on the constrained fit is $50\%$ of its central 
value. We have done a similar study for the first excited state of the 
$\eta$ meson using the lattice data from Ref.~\cite{Dudek:2013yja}. The 
results are shown in the right panel of Fig.~\ref{fig:mesonresonances}. 
We again observe that the $\sigma$ term obtained from the unconstrained 
fit lies within a variation of $50\%$ of the value of $\sigma=15(3)$ 
MeV obtained from the constrained fit. 

In our fit estimates we could not observe any significant difference of the 
$\sigma$ term of the excited states of mesons in different spin, parity sectors 
from their ground state values within error. For example the excited states of 
both strange and nonstrange isoscalars have $\sigma=10(3),~33(5)~\text{ MeV}$, 
also consistent with the values corresponding to their ground states. Hence 
we have taken the $\sigma$ terms of the excited states to be the same as the 
ground state mesons in different spin, parity sectors but with a $50\%$ error 
to take into account the systematic uncertainties.

\begin{figure}[h!]
\includegraphics[width=0.45\textwidth]{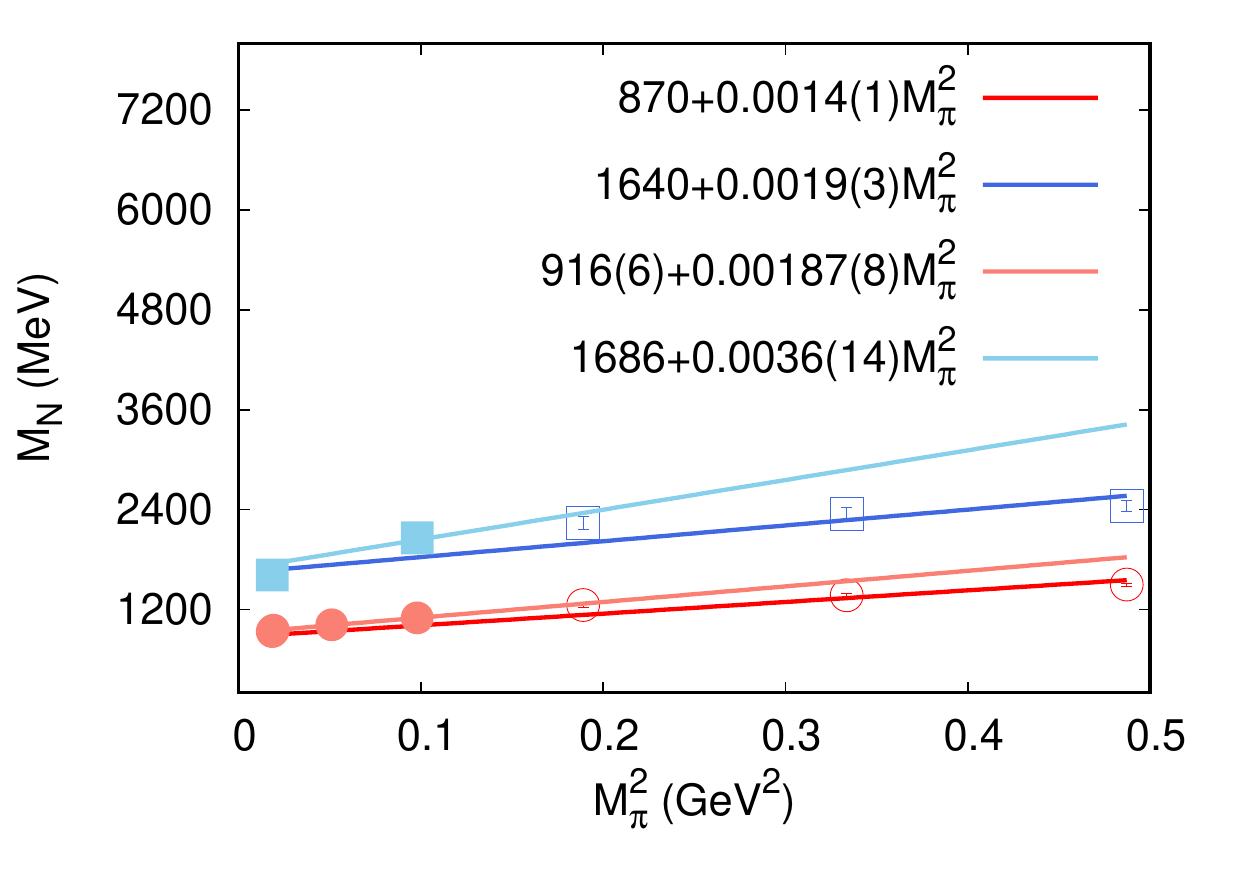} 
\caption{The fits to the mass of the ground state (shades of red) and the 
first excited state (shades of blue) nucleon as a function of the square 
of the pion mass, using the $2+1$ flavor data (open symbols) from Ref.~\cite{Virgili:2019shg} 
and $2+1+1$ flavor data (closed symbols) from Ref.~\cite{Mondal:2020cmt}.
All fit results are in units of MeV.}
\label{fig:nucleonsigma} 
\end{figure}

In the baryon sector, we could find the data for the mass of the ground 
and excited states as a function of the pion mass only for the nucleon 
(octet) both for $2+1$~\cite{Virgili:2019shg} and $2+1+1$ 
flavor~\cite{Mondal:2020cmt} QCD, which are shown in 
Fig.~\ref{fig:nucleonsigma}. For the $2+1+1$ flavor data which 
correspond to pion masses of $313, 225.9, 135.6$ MeV respectively a fit 
to the ground state yields $\sigma=36(2)~\text{MeV}$, whereas for the 
excited state the data are available only for the first and last 
values of the pion mass stated above, which gives a 
$\sigma=68(27)~\text{MeV}$. For the $2+1$ flavor data, a fit to the 
positive parity nucleon mass data and its first excited state gives 
$\sigma=27\pm 2$ and $36\pm 6~\text{ MeV}$, respectively, which are 
quite close to each other. For the $2+1$ flavor fits we assumed  that 
the ground state mass of the nucleon is reduced by $70$ MeV in the 
chiral limit and the difference of its mass with the first excited 
state remains similar, $\approx 770$ MeV, as seen in 
Ref.~\cite{Copeland:2021qni}. A more sophisticated fit strategy 
discussed in Ref.~\cite{Copeland:2021qni} gives a ground state octet 
$\sigma \approx 44 ~\text{MeV}$, which we use in our paper. We 
nevertheless performed these simple fits to understand how much the 
$\sigma$ terms of the excited state deviate from the ground state of 
the baryon octet. We find that the central value varies at most by 
$1.5$ times from the ground state estimates but agree with each other 
within errors. We summarize our findings of our fits in the 
Table~\ref{table:octetdecupsigma}.

\begin{table}[]
\caption{\label{table:octetdecupsigma} Table summarizing the $\sigma$ terms 
for the ground state and excited states of several mesons and the nucleon from 
the fits to the lattice data we performed. }
\begin{tabular}{|c||c|c|c|c|c|c|c|c|}
\hline
Hadron state & $\pi$ & $\eta$ & $\eta^{'}$ & $\rho$ & $\omega$ &  $\phi$ & $K^*$ & N \\
\hline
\hline
Ground state $\sigma (\text{MeV})$ &-&$10\pm 7$ & $9.5\pm 4.5$ &  $27\pm 4$ & $12\pm 2$ & $0.3 \pm 0.3$ & $10\pm 1$ & -\\
\hline
First excited state $\sigma (\text{MeV})$ & $26\pm 3$ & $15\pm 3$ &-& - & $33\pm 5$ &  $10 \pm 3$ & -& $68\pm 27$ \\
\hline
\end{tabular}
\end{table}
We have also roughly estimated the $\sigma$ terms for the higher spin 
excited states of the baryon octet and decuplet using the data 
from~\cite{Edwards:2011jj}. Here again we have the mass data for the 
excited states for two different pion masses $396$ and $524$ MeV, 
respectively. For the spin $\frac{1}{2}, \frac{3}{2}, \frac{5}{2}$, and 
$\frac{7}{2}$ first excited nucleon states we have $\sigma=37, 51, 51,$ 
and $51~\text{ MeV}$, respectively by performing a two-term 
unconstrained fit. For the first excited state of $\Delta$ we 
have $\sigma=40, 39$, and $38~\text{ MeV}$, for spins  $\frac{1}{2}, 
\frac{3}{2}$, and $\frac{5}{2}$, respectively. Since there are only two 
data points in each fit, we could not estimate the errors of the $\sigma$ 
terms but they all lie within $1 \sigma$ error of the ground state estimates~\cite{Copeland:2021qni}. Henceforth in this work the $\sigma$ 
terms for these excited states of baryons are also taken to be the same 
as their ground states.  Similarly to our prescription for the excited 
states of mesons we have considered a $50\%$ variation of the central 
values of these $\sigma$ terms when estimating the error on the chiral 
condensate.

\section{RELATIVE CONTRIBUTIONS TO THE RENORMALIZED CONDENSATE FROM MESONS AND BARYONS}
\label{appdx:b}
\begin{figure}[h]
\includegraphics[width=0.3\textwidth]{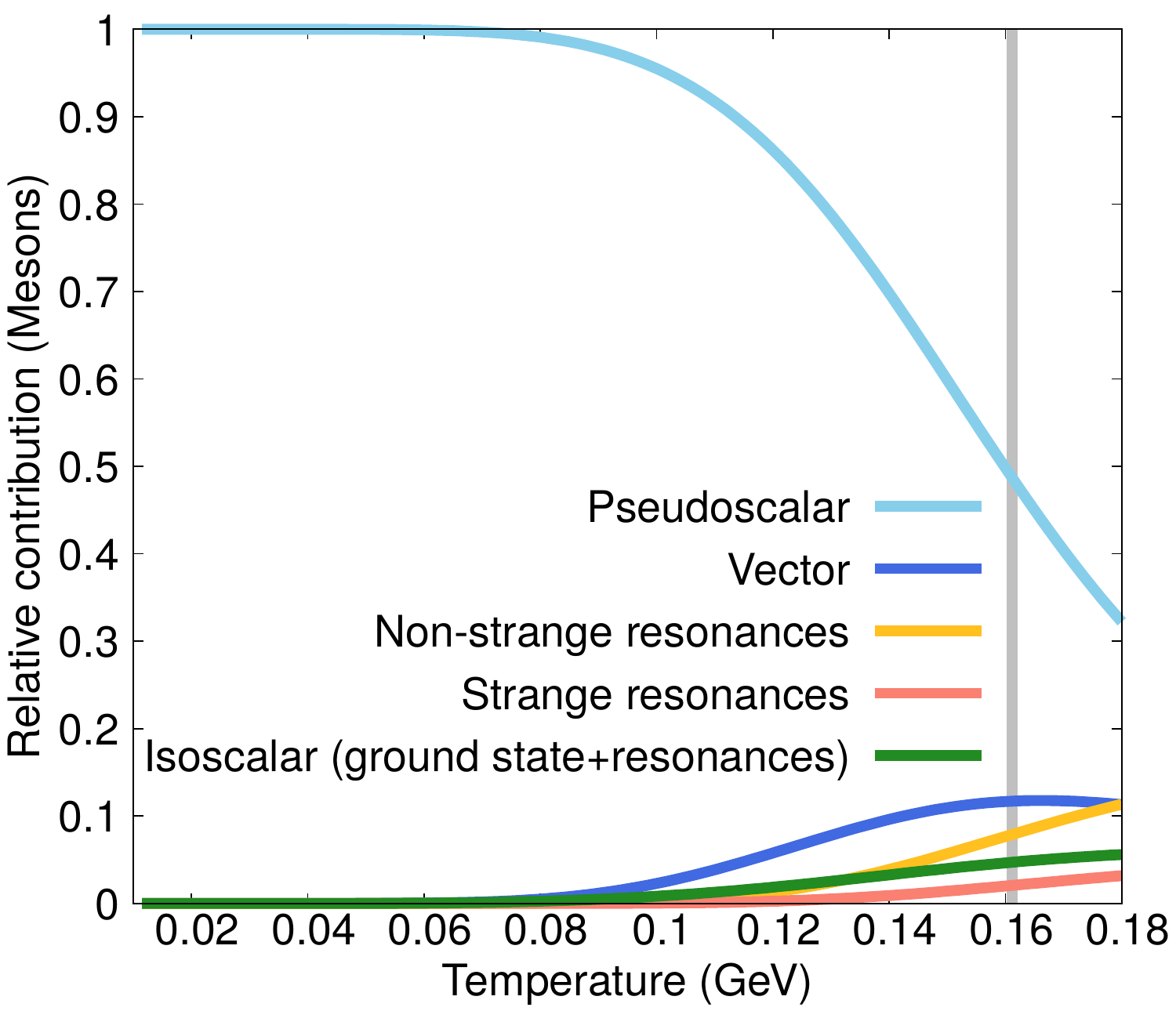}
\includegraphics[width=0.3\textwidth]{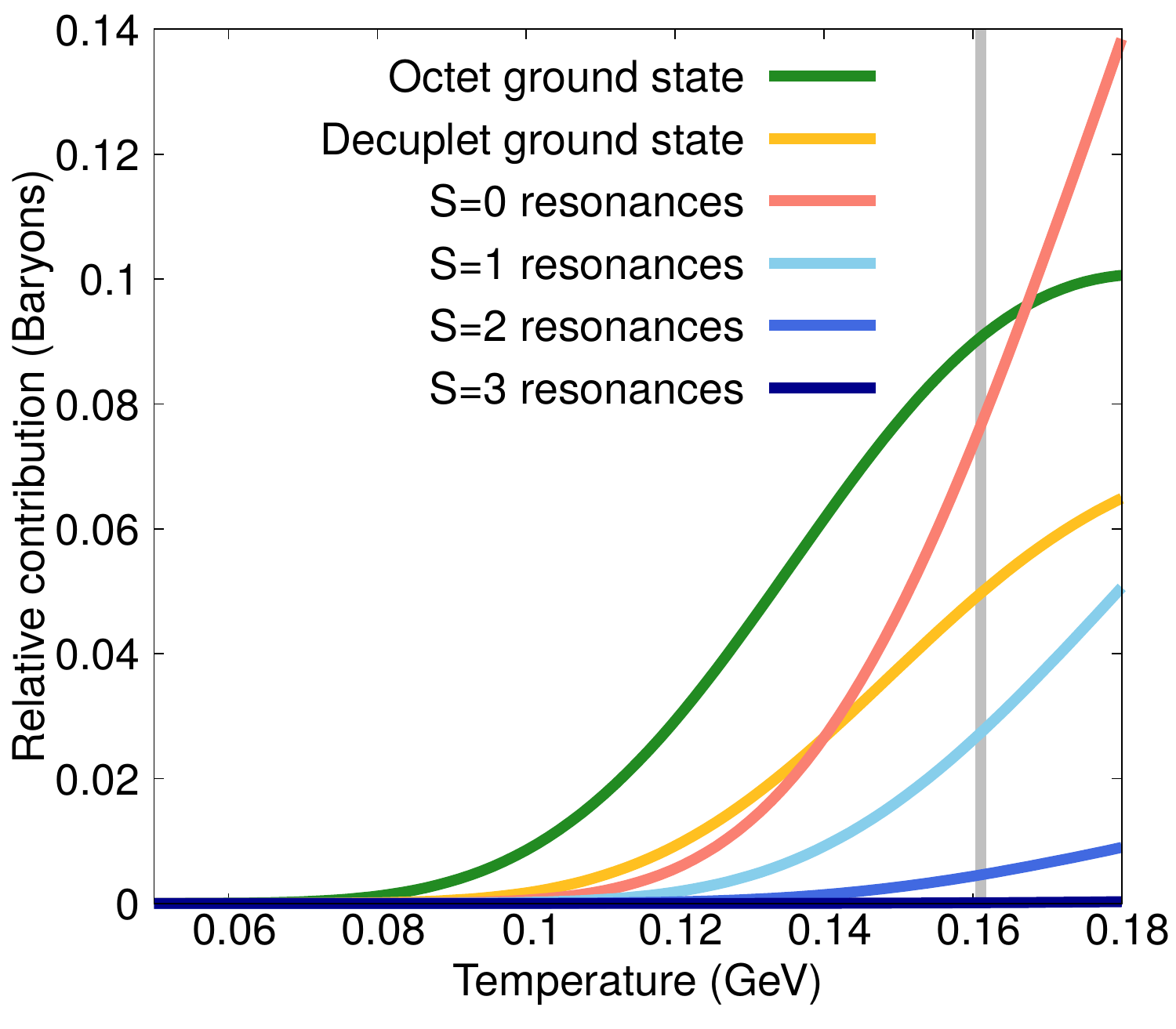} 
\includegraphics[width=0.3\textwidth]{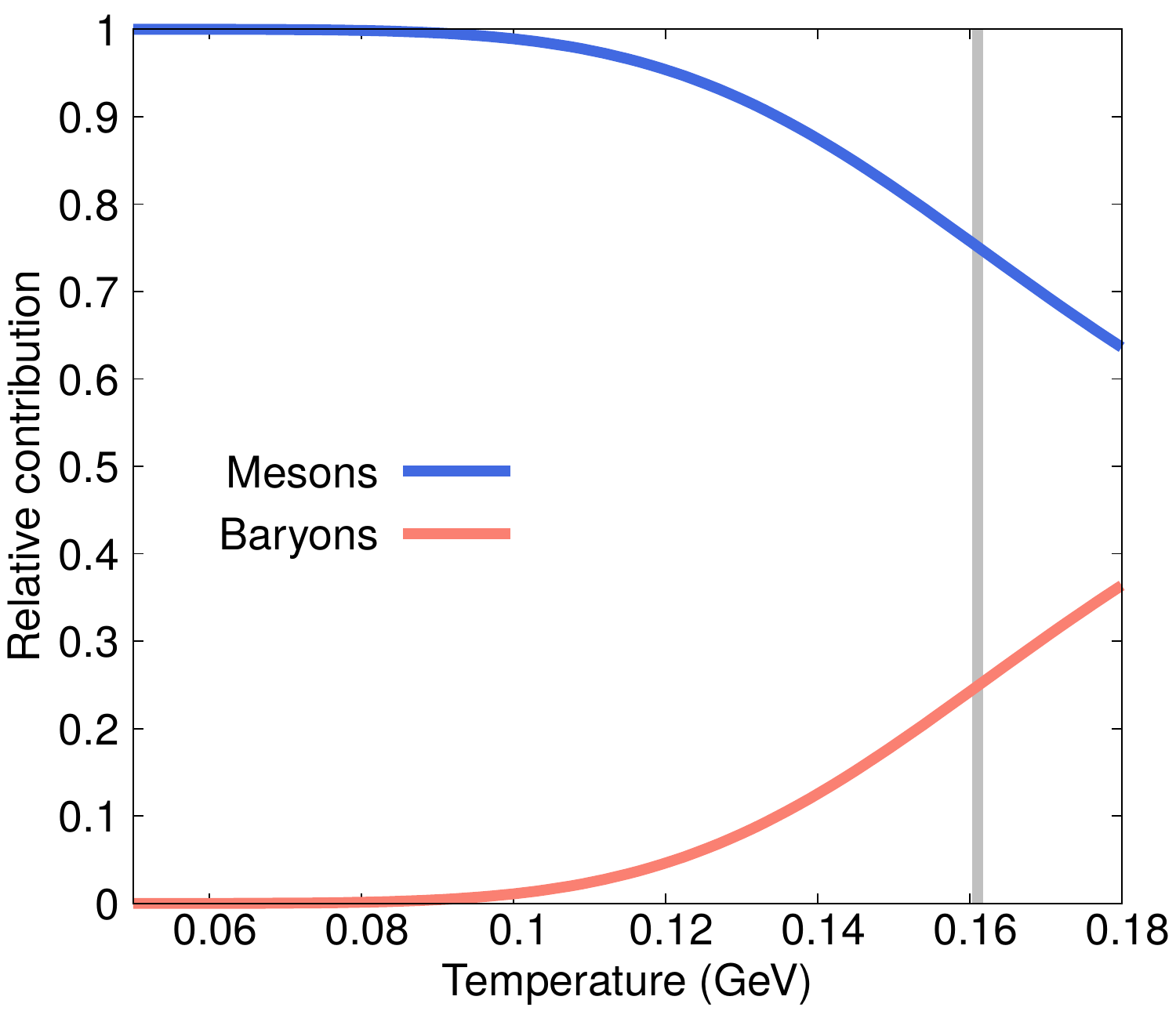} 
\caption{The right panel shows that the relative contribution to the 
renormalized chiral condensate comes from mesons ($80\%$) followed 
by baryons ($20\%$). The left and central panels show the relative 
contribution to the meson and baryon sectors respectively due to 
different states.}
\label{fig:rel}
\end{figure}

We discuss here the relative contributions of the mesons and baryons with 
different quantum numbers to the renormalized chiral condensate as a function 
of temperature. The relative contributions of different hadrons are shown in 
Fig.~\ref{fig:rel}.  Within the meson sector, we find that the largest relative 
contribution to the renormalized chiral condensate near the crossover transition 
temperature comes from the ground state pions and kaons followed by the vector 
meson. The combined relative contribution from the excited states of both strange 
and nonstrange mesons is only $1/3$ of the total ground state contributions. The 
overall meson contributions to the renormalized chiral condensate is almost $75\%$ 
at a temperature $T=160~$ MeV. The remaining contributions come from the 
baryon sector, dominantly from the ground state octet and the excited states, 
primarily those of the nucleons.
\end{appendix}
\bibliography{references_paper1.bib}

\end{document}